\documentclass[aps,prb,twocolumn,amsmath,amssymb,groupedaddress,showpacs]{revtex4}
\pdfoutput=1

\usepackage{graphicx}% Include figure files
\usepackage{dcolumn}% Align table columns on decimal point
\usepackage{bm}% bold math
\usepackage{caption}
\usepackage{subcaption}
\usepackage{color}
\usepackage{float}

\usepackage{ulem}
\usepackage{xcolor}

\usepackage{xcolor,soul}

\begin{document}

%\preprint{APS/123-QED}

\title{Continuous degeneracy of the fcc kagome lattice with magnetic dipolar interactions}% Force line breaks with \\

\author{A. R. Way}
%  \altaffiliation[Also at ]{Physics Department, XYZ University.}%Lines break automatically or can be forced with \\
\author{K. P. W. Hall}%
%  \email{Second.Author@institution.edu}
\author{I. Saika-Voivod}
\author{M. L. Plumer}
\affiliation{%
Department of Physics and Physical Oceanography, Memorial University of Newfoundland, St. John's, Newfoundland, Canada A1B 3X7
}%

\author{B. W. Southern}
%  \homepage{http://www.Second.institution.edu/~Charlie.Author}
\affiliation{
 Department of Physics and Astronomy, University of Manitoba, Winnipeg, Manitoba, Canada R3T 2N2
}%

% \affiliation{%
%  Authors' institution and/or address\\
%  This line break forced with \textbackslash\textbackslash
% }%

% \collaboration{CLEO Collaboration}%\noaffiliation

\date{\today}% It is always \today, today,

\begin{abstract}
Results are presented on analytic and computational analyses of the spin states associated with a 3D fcc lattice composed of ABC stacked kagome planes of magnetic ions with only long-range
dipole-dipole interactions. Extending previous work on the 2D kagome system, where discrete
six-fold discrete degeneracy of the ground state was revealed [Holden {\it et al.} Phys. Rev. B {\bf 91}, 224425 (2015)], we show that the
3D lattice exhibits a continuous degeneracy characterized by just two spherical angles involving six
sublattice spin vectors. Application of a Heat Bath Monte Carlo algorithm shows that thermal fluctuations 
reduce this degeneracy at very low temperature in an order-by-disorder process. A
magnetic field applied along directions of high symmetry also results in lifting the continuous degeneracy to a subset of states from the original set of ground states. 
Metropolis Monte Carlo simulation results are also presented on the temperature and system size dependence of the energy, specific heat, and magnetization, providing evidence for a phase transition at T $\simeq$ 0.38 (in units of the dipole strength). The results
can be relevant to a class of magnetic compounds having the AuCu$_3$ crystal structure.
%\begin{description}
%\end{description}
\end{abstract}

\maketitle

%\tableofcontents

%\input{intro}
\section{Introduction}\label{sec:intro}

Geometrically frustrated spin systems are typically associated with short-range Heisenberg-like antiferromagnetic exchange interactions.\cite{frustration}
A more subtle geometry-induced frustration can also occur with only dipole coupling between spin vectors.  Features
which lead to this frustration include an antiferromagentic (AF)-like contribution to the dipole interaction, the first term in Eq. (1) (where D is the dipole strength),
as well as the second term which involves explicit coupling of spin and lattice vectors.  The long-range nature of the dipole coupling 
gives additional aspects (equivalent to a decaying higher-neighbor exchange in the first term) which are usually revealed only through numerical 
calculations. We adopt the dipole interaction of the form
\begin{equation}\label{eq:DipoleHamiltonian}
    E_{dip} = D\sum_{i < j} \left[ \dfrac{\left( \vec{S}_{i} \cdot \vec{S}_{j} \right)}{r_{ij}^{3}} - 3 \dfrac{\left( \vec{S}_{i} \cdot \vec{r}_{ij} \right) \left( \vec{S}_{j} \cdot \vec{r}_{ij} \right)}{r_{ij}^{5}} \right]
\end{equation}
where the dipole strength is given by $D=\frac{\mu_0}{4 \pi}\frac{(g\mu_B S)^2}{a^3}$ with $a$ being the near neighbour lattice
spacing, $\vec{r}_{ij}$ is the dimensionless vector connecting sites $i$ and $j$,
and $S_i$ is a unit spin vector. For convenience,  all calculated quantities involving the energy and temperature have units relative to the dipole strength.    
In the present work, exchange terms are omitted in an effort to explore fully the implications of the long-range effects of the 3D frustrated lattice structure described below, as well as having a model applicable to artificial nanostructures.\cite{artificial} Spin structures without and with an applied magnetic field, as well as at finite temperature, are explored in the case of a 3D fcc lattice composed of ABC stacked kagome planes along cubic $\langle 111 \rangle$ directions with classical spin vectors interacting only through the dipole coupling energy Eq. (1). The structure is inspired by magnetic compounds, such as IrMn$_3$,\cite{tomeno,kren} which adopt a AuCu$_3$ crystal structure, as shown in Fig. 1.  Unlike the regular fcc lattice, the impact of the inherent AF frustration plays a more important role in the dipolar spin states of the fcc kagome lattice. 

\begin{figure}[H]
       	\centering
        \includegraphics[width=0.9\columnwidth, keepaspectratio, trim = 2.5cm 5cm 2cm 2.5cm, clip]{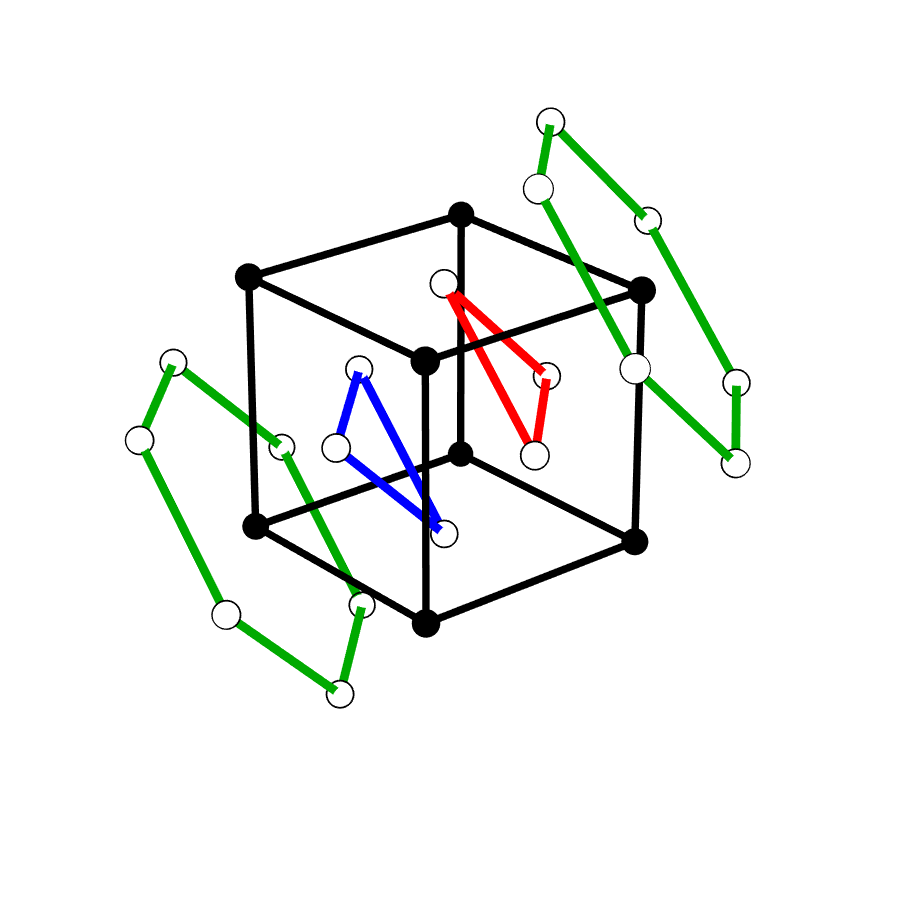}
        \caption{(Colour online) The fcc kagome lattice based on a generic AB$_3$ compound having the AuCu$_3$ crystal structure with B-site magnetic ions (open circles) on the cube faces forming stacked 2D kagome layers along  $\langle 111 \rangle$ axes. Non-magnetic A-ions (filled circles) are at the corners of the unit cell. (Adapted from Ref. [\onlinecite{leblanc1}]).}
        \label{fig:ReducedGS}
\end{figure}

In many cases, a consequence of geometrical frustration is  a degeneracy of spin configurations in the ground state, normally when only nearest-neighbor (NN) antiferromagnetic exchange interactions are included.
This is particularly relevant for the 2D kagome lattice composed of corner-sharing triangles where the degeneracy is macroscopic, involving all the structures with the basic 120$^\circ$ spin configuration around each triangle.\cite{mekata2003,harris1992,zhito2008,schnabel2012,chern2013,taillef2014}  In the case of only dipole coupling between spins on the 2D kagome lattice, this degeneracy is reduced to six-fold due to the direct spin-lattice coupling.\cite{maksy2015,holden2015,maksy2017}
In general, the effect of thermal fluctuations or an applied magnetic field on frustrated spin systems is to reduce the degeneracy.  For the kagome lattice with NN antiferromagnetic exchange, thermal fluctuations select co-planar q=0  spin structures at very low temperatures. With only dipole coupling, the six-fold (in-plane) spin degeneracy is largely unaffected by temperature except on cooling below T$_C$ $\simeq$ 0.43 where the system tends to lock-in to one (or more) of the six spin configurations at T $\simeq$ 0.2. At all temperatures below T$_C$ the 2D kagome dipole system displays a net magnetization.  

Spatial dimensionality of the lattice also plays an important role on both the ground state degeneracy and impact of thermal fluctuations.  
Previous studies of the 3D fcc kagome lattice with NN antiferromagnetic exchange J (involving 8 neighboring sites), for both XY and Heisenberg models,  showed that it also exhibits 120$^{\circ}$ spin configurations associated with q=0 order as in the 2D case (but with reduced degeneracy), and a phase transition to long-range order at T$_{N}$ = 0.760J and T$_{N}$ = 0.476J, respectively.\cite{hemmati,leblanc1,leblanc2}
In the case of the triangular lattice with only dipole interactions, the ground state is ferromagnetic in 2D as well as for the 3D hexagonal structure.\cite{mckeehan,tomita,johnston}
The impact of higher dimensionality on ground state dipolar spin structures in going from the 2D square lattice to 3D cubic structures is more complicated.  The dipolar 2D square lattice exhibits AF order\cite{debell,johnston} whereas the regular fcc lattice shows ferromagentic order.\cite{bouchard,johnston} In contrast, and of particular relevance to the present work, the simple cubic dipolar lattice is
characterized by a four sublattice spin configuration with continuous degeneracy involving two angles,\cite{belobrov} also seen in cubic clusters.\cite{schonke,kure}  Monte Carlo simulations on the cubic system suggest a discontinuous phase transition at T$_C$ $\simeq$ 0.56.\cite{romano} Interest in the interplay between frustration and dipole interactions has been enhanced by the discovery of spin ice materials.\cite{spinice}

In this work we examine ground-state spin structures of the fcc kagome dipolar lattice using an Effective Field Method\cite{walker} (EFM) with lattice sums performed using Ewald techniques.\cite{ewald}  The results reveal degenerate ground-state configurations characterized by 
six sublattice spins, composed of three spins per adjacent (111) kagome plane, with antiferromagnetic alignment between planes, yielding a zero net magnetic moment (in contrast with the 2D case).  The six spin vectors are characterized by just two angles.  This continuous degeneracy is shown to be removed by an order-by-disorder\cite{obd} process using a Heat Bath algorithm.  Similar degeneracy reduction is also shown to be achieved with magnetic field cycling.  Finally, Metropolis Monte Carlo (MC) simulations are used to demonstrate a phase transition to long-range magnetic order at T$_C$ $\simeq$ 0.38 with a lock-in transition at T $\simeq$ 0.3. 

The remainder of this work is organized as follows. We describe the model used in Sec.~\ref{sec:groundstate} as well as the characterization of the ground state. The selection of particular ground states by thermal fluctuations at very low temeprature is described in Sec.~\ref{sec:finitetemp}. The impact of an applied magnetic field is described in \ref{sec:appliedfield}. MC simulation results on various thermodynamic quatities are presented in Sec.~\ref{sec:mc}, with a summary and conclusions given in Sec.~\ref{sec:summary}.

\section{Ground State Characterization}\label{sec:groundstate}

\begin{figure}[H]
        \centering
        \includegraphics[width=1\columnwidth, keepaspectratio]{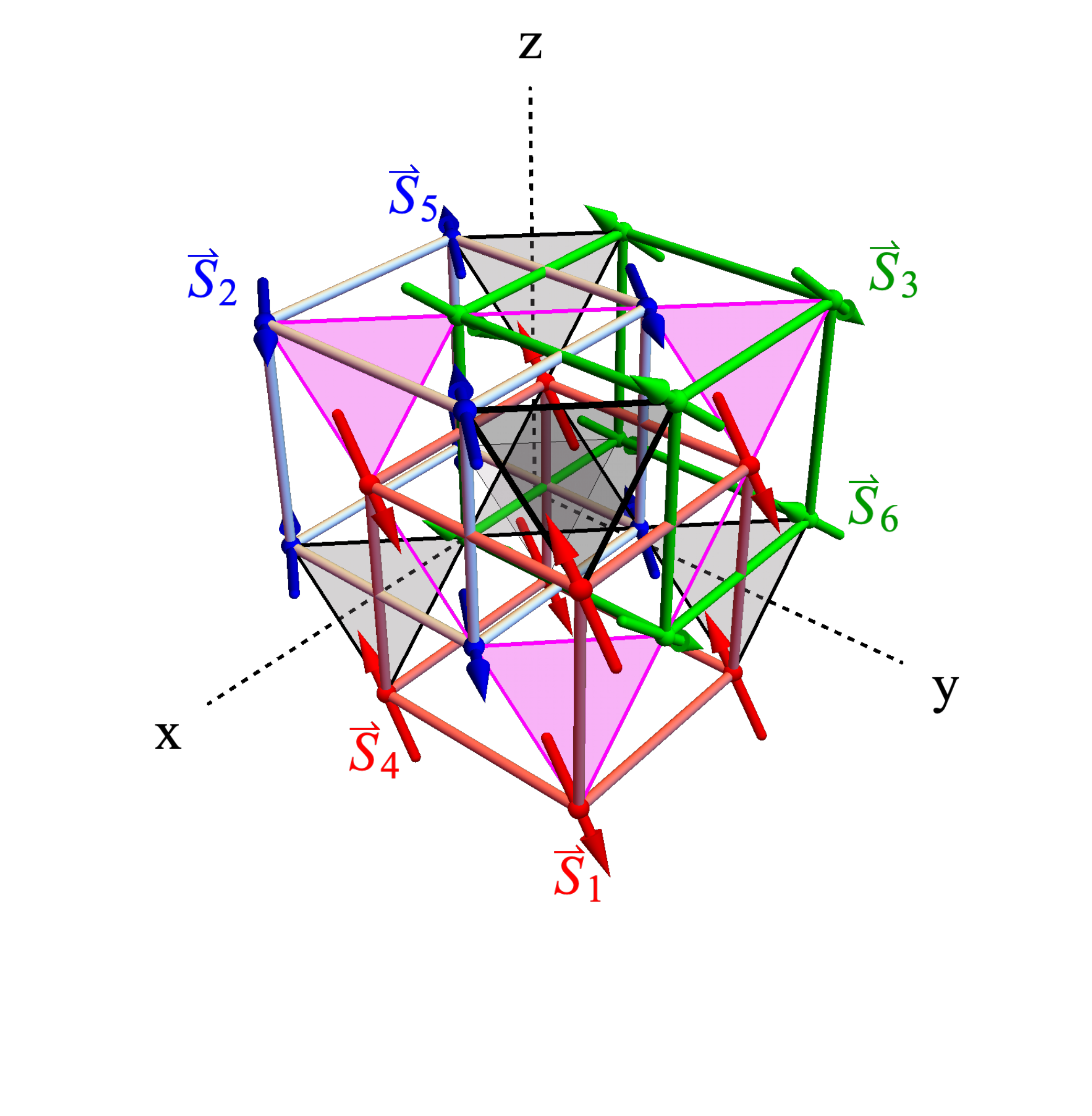}
        \caption{The 3D fcc lattice with six sublattice spin vectors on three interpenetrating cubic lattices.  In our labelling convention, $\vec{\mathbf{S}}_1$ (two of which are in the $z=0$ plane) and $\vec{\mathbf{S}}_4 = -\vec{\mathbf{S}}_1$ are red, $\vec{\mathbf{S}}_2$ (two of which are in the $y=0$ plane) and $\vec{\mathbf{S}}_5 = -\vec{\mathbf{S}}_2$ are blue, and $\vec{\mathbf{S}}_3$ (two of which are in the $x=0$ plane) and $\vec{\mathbf{S}}_6 = -\vec{\mathbf{S}}_3$  are green.  The directions of the spins are those for a ground state with $\theta=140^\circ$ and $\phi=80^\circ$.  The view is nearly down the $\left[111\right]$ axis, with the triangle with the darkest outline belonging to the kagome plane closest to the eye.}
        \label{fig:3dfcc}
\end{figure}

EFM simulations were performed on a three-dimensional lattice consisting of $L$ ABC stacked 2D kagome planes (along cubic $\langle 111 \rangle$ axes), with each plane occupied by $\dfrac{3}{4}$($L \times L$) unit spin vectors\cite{hemmati,leblanc1,leblanc2} at  lattice sizes $L =$ 6, 12, and 18 using periodic boundary conditions.  Several thousand runs were executed with random initial spin configurations to determine the lowest energy state, where each run involved several hundred sweeps through the lattice.
Analysis of the results reveals a many-fold degenerate ground state allowing configurations with a six sublattice spin structure, having three sublattice spins per kagome layer, alternating in sign along the $\langle 111 \rangle$ axes (i.e., adjacent layers are AF aligned). The  basis spin vectors themselves can also be viewed as occupying the sites of three simple cubic lattices imbedded in the fcc kagome structure, as shown in Fig. \ref{fig:3dfcc}.  

By inspection of numerous resulting ground-state spin vectors, it was determined that every configuration is found to be characterized by the following set of equations:
\begin{equation}
\begin{split}
S_{1x} = & \sin{\theta} \cos{\phi}, \\
S_{1y} = & \sin{\theta} \sin{\phi}, \\
S_{1z} = & \cos{\theta}, \\
\alpha = & \frac{(2S_{1x}^2-1)}{2 S_{1z}}, \\
\beta = & \pm \sqrt{1-S_{1x}^2-\alpha^2}.
\end{split}
\label{eqn:spinparameters}
\end{equation}
These equations act as elementary building blocks for the components of the six ground state spin vectors.  There are only two parameters that characterize the resulting ground states,  $\theta$ and $\phi$, which are the polar and azimuthal angles of  spin ``1'' ($\vec{\mathbf{S}}_1$), where the polar axis lies along the positive z axis [001] of a Cartesian reference frame. The values of $\theta$ and $\phi$  are restricted to ensure real values of $\beta$ (note that the Heisenberg model is also characterized by only two spin vector angles without any restriction on the values the angles may take.) Only six spins are required to fully characterize the ground states. The six ground state spin vectors themselves may be constructed as follows:
\begin{equation}
\begin{split}
\vec{\mathbf{S}}_1 = & (S_{1x}, S_{1y}, S_{1z}), \\
\vec{\mathbf{S}}_2 = & (\alpha, \beta, -S_{1x}), \\
\vec{\mathbf{S}}_3 = & (-\beta, -S_{1z}-\alpha,  -S_{1y}), \\
\vec{\mathbf{S}}_4 = & -\vec{\mathbf{S}}_1, \\
\vec{\mathbf{S}}_5 = & -\vec{\mathbf{S}}_2, \\
\vec{\mathbf{S}}_6 = & -\vec{\mathbf{S}}_3.  \\
\end{split}
\label{eqn:spinvectors}
\end{equation}
Here, $\vec{\mathbf{S}}_1$, $\vec{\mathbf{S}}_2$, and $\vec{\mathbf{S}}_3$ are the three spins on a given triangle in one (111) kagome plane and $\vec{\mathbf{S}}_4$, $\vec{\mathbf{S}}_5$, and $\vec{\mathbf{S}}_6$ are spins on a triangle above or below on an adjacent (111) kagome plane.
 In all figures, $\vec{\mathbf{S}}_1$ appears in red, $\vec{\mathbf{S}}_2$ appears in blue, and $\vec{\mathbf{S}}_3$ appears in green.

We note that the relations between components of $\vec{\mathbf{S}}_1$, $\vec{\mathbf{S}}_2$, and $\vec{\mathbf{S}}_3$ can be obtained by directly calculating the system's dipolar energy with a finite radial cutoff as a function of the components of the three independent spins, and assuming the six-sublattice structure given in Eq. ~(\ref{eqn:spinvectors}).  
Consider $\vec{\mathbf{S}}_1$ in Fig.~\ref{fig:3dfcc} as a central spin.  It resides on a vertex in the red simple cubic lattice.  The dipolar energy is zero for this sublattice (this is straight forward to show for a given spin and its six nearest neighbours).  The central $\vec{\mathbf{S}}_1$ spin also resides in the middle of a face of the blue simple cubic lattice of alternating $\vec{\mathbf{S}}_2$ and $-\vec{\mathbf{S}}_2$ spins, and similarly for the green simple cubic lattice of alternating $\vec{\mathbf{S}}_3$ and $-\vec{\mathbf{S}}_3$ spins.  On the level of nearest neighbours, $\vec{\mathbf{S}}_1$ is in the middle of two orthogonal squares, one blue and one green.  Because neighbouring spins alternate signs, the AF-like contribution to the energy is zero (the first term in the sum in Eq.~(\ref{eq:DipoleHamiltonian})).  This is in contrast to the 2D kagome dipolar system where the AF-like contribution from nearest neighbours is not zero, and hence the sublattice structure of the ground state emerges only after considering next-nearest neighbours.\cite{holden2015}  Thus, only the lattice-coupling term of Eq.~(\ref{eq:DipoleHamiltonian}) contributes to the ground state energy in the 3D kagome system.  In terms of independent spin components of $\vec{\mathbf{S}}_1$, $\vec{\mathbf{S}}_2$, and $\vec{\mathbf{S}}_3$  the per particle dipolar energy of the six-spin system is,
\begin{eqnarray}
%u_D = \delta ( S_{1x}S_{3z} &+& S_{1z}S_{3x} + S_{1y} S_{2z}   \\
%                    &+& S_{1z} S_{2y}  + S_{2y}S_{3x} + S_{2x}S_{3y} ) D,  \\ \nonumber
u_D = \delta ( S_{1x}S_{2z} &+& S_{2x} S_{1z} + S_{1y} S_{3z}   \\
                    &+& S_{3y} S_{1z}   + S_{2x} S_{3y} + S_{3x}S_{2y} ) D,   \nonumber
\end{eqnarray}
where the nearest neighbour distance is unity and $\delta$ is a factor that starts at a value of 2 when only considering nearest neighbours and decreases with increasing the radial cut-off distance (to 1.7030969 when the cutoff is 22). 
A minimization of this energy with respect to the spin components subject to normalization of the spin vectors (through the method of Lagrange multipliers, for example) yields the expressions for 
$\vec{\mathbf{S}}_1$, $\vec{\mathbf{S}}_2$, and $\vec{\mathbf{S}}_3$ in Eqs.~\ref{eqn:spinparameters}-\ref{eqn:spinvectors}, and a minimum value of the quantity in brackets of $-3/2$.  This ground state energy of $-3\delta/2 =  -2.55465$ compares well with our Ewald result of -2.55458.

%With this sublattice structure, any one of $\vec{\mathbf{S}}_1$, $\vec{\mathbf{S}}_2$, and $\vec{\mathbf{S}}_3$ taken as a central spin can be viewed as a vertex in a simple cubic lattice of the type shown with a single color in Fig~\ref{fig:3dfcc}, for which the dipolar energy is zero, and as a spin in the centre of a face in each of two simple cubic lattices each formed from the other two spins, i.e. the other two colored cubes in Fig.~\ref{fig:3dfcc}.  

\begin{figure}[H]   % Should a reduced ground state be introduced here or later?
    \centering
    \includegraphics[width=1\columnwidth, keepaspectratio]{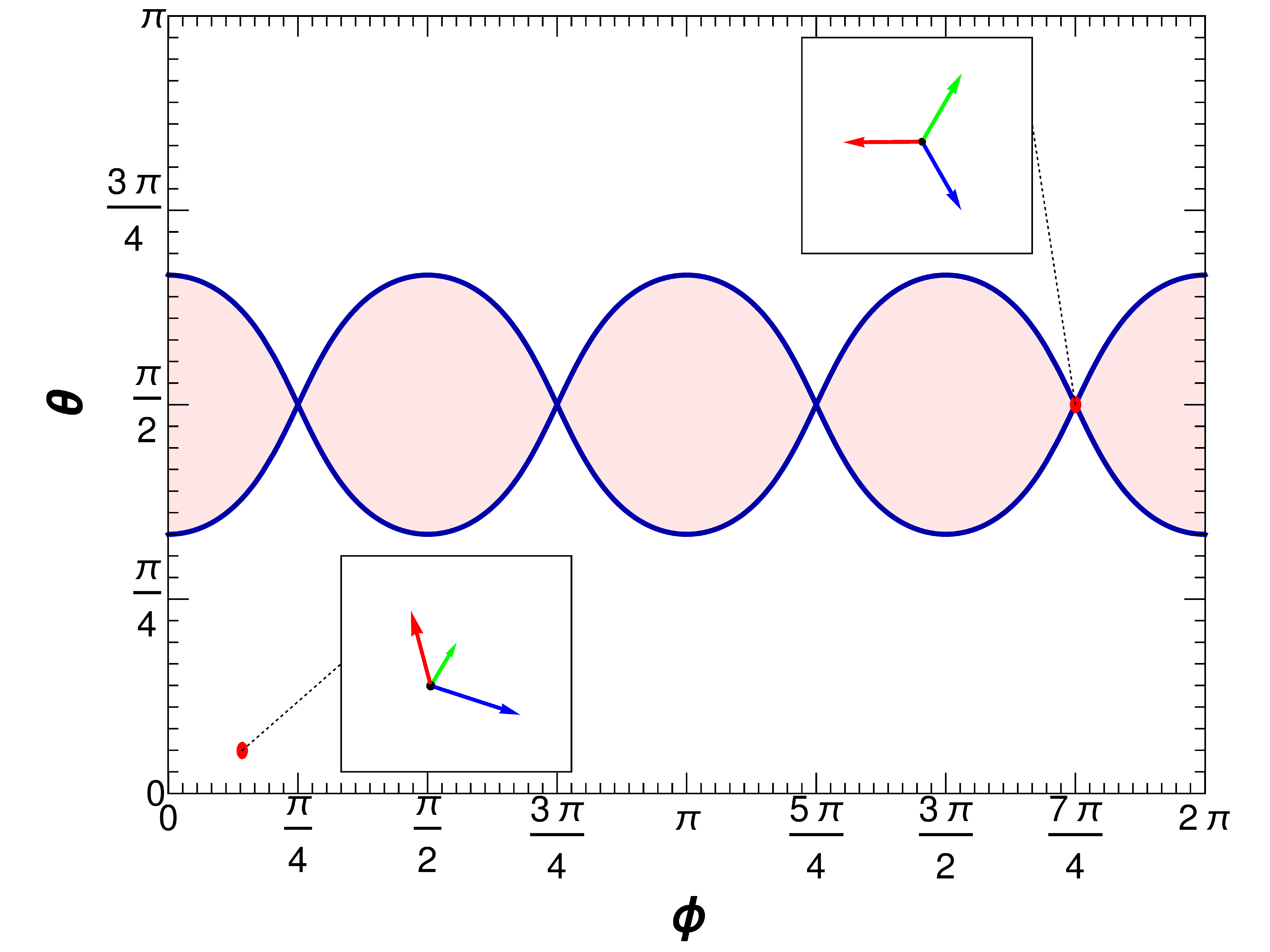}
    \caption{ (Colour online) The 2-dimensional space describing the continuous degeneracy of the fcc kagome ground state. Example states are represented in the insets in the image. Both insets are viewed in the $[ \bar{1}\bar{1}\bar{1} ]$ direction. Within the shaded area enclosed by the curves the values of $\theta$ and $\phi$ do not yield valid states.}
    \label{fig:degenplanefull}
\end{figure}

Any ${(\theta, \phi)}$ pair chosen from the plane in Fig.~\ref{fig:degenplanefull} will give rise to a valid ground state of the same energy with the exception of those pairs of ${(\theta, \phi)}$ that lie within the shaded region bounded by the blue curve of the graph. Within this bound region, $\beta$ in Eq.~\ref{eqn:spinparameters} has an imaginary part and thus spin states characterized by those angle pairs ($\theta,\phi$) are not allowed and do not appear in the EFM results. At each node of each bounded area, $\alpha$ in Eq.~\ref{eqn:spinparameters} becomes undefined,
but the limit as $\theta$ approaches $\pi/2$ for $\phi=\pi/4$, $3\pi/4$, $\dots$ is well defined. 

Note that the pattern in Fig.~\ref{fig:degenplanefull} repeats in $\phi$ with a period of $\frac{\pi}{2}$ due to crystal symmetry considerations. Nodes occur at $\phi = \frac{2n+1}{4 \pi}$ where $n$ is an integer. The plane is simplified further due to reflections about the axes in the middle of each bounded region. The three spins at node states lie in a $\{111\}$ plane.  For example, for $\theta = \pi/2$ and $\phi = \pi/4$, 

\begin{equation}
\begin{split}
\vec{\mathbf{S}}_1 = & (\frac{1}{\sqrt{2}},\frac{1}{\sqrt{2}}, 0), \\
 \vec{\mathbf{S}}_2  = & (0,\frac{1}{\sqrt{2}},-\frac{1}{\sqrt{2}}), \\
 \vec{\mathbf{S}}_3 = & (-\frac{1}{\sqrt{2}},0,-\frac{1}{\sqrt{2}}), \\
\end{split}
\label{eqn:nodevectors}
\end{equation}
are in the $(\bar{1},1,1)$  plane.

%The reduced plane that contains all pairs of $\theta$ and $\phi$ that give rise to a dipolar ground state is shown in Fig.~\ref{fig:degenplane}, where valid $\theta$-$\phi$ pairs lie above the boundary shown.

%\begin{figure}[ht]
%	\includegraphics[width=0.43\textwidth, keepaspectratio]{img_Andrew/degeneracyplane.png}
%	\caption{One section (0, $\frac{\pi}{4}$) of the original degeneracy plane that is equivalent to all other sections of the plane due to symmetry operations.}
%	\label{fig:degenplane}
%\end{figure}

Analysis of the relations Eq.~\ref{eqn:spinparameters} reveals that the region of allowed $(\theta, \phi)$ pairs satisfy,
\begin{align}
\label{eqn:bound}
\pi \geq \theta \geq \left\{ \begin{array}{cc} 
                \frac{2\pi}{3} & \hspace{5mm} \phi=0 \\
                \pi - \sin^{-1}{\left(\sqrt{ \frac{4-\sqrt{16-6(1-\cos{(4\phi)})}}{1-\cos{(4\phi)}} }\right)} & 0 < \phi \leq  \frac{\pi}{4} \\
                \end{array} \right.
\end{align}
Thus, the entirety of the ground state is characterized by Eqs.~(\ref{eqn:spinparameters})-(\ref{eqn:spinvectors}) and the parameters ${(\theta, \, \phi)}$ such that they satisfy Eq.~(\ref{eqn:bound}).

\begin{figure}[H]
	%TODO Change this image with a better one
	\centering
	\includegraphics[width=0.6\columnwidth, keepaspectratio]{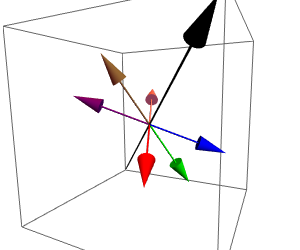}
	\caption{The six sublattice spins corresponding to a non-planar state conjoined at their ends for clarity and illustration. The [111] axis is indicated by the long black vector.}
	\label{fig:sampgs}
\end{figure}

By generating the spin vectors described by Eqs.~(\ref{eqn:spinparameters})-(\ref{eqn:spinvectors}), spin configurations  
from what we will call planar and non-planar states can be obtained. An example of
a non-planar state is shown in Fig. \ref{fig:sampgs}.

\begin{figure}[H] 
    \centering
    \includegraphics[width=1\columnwidth]{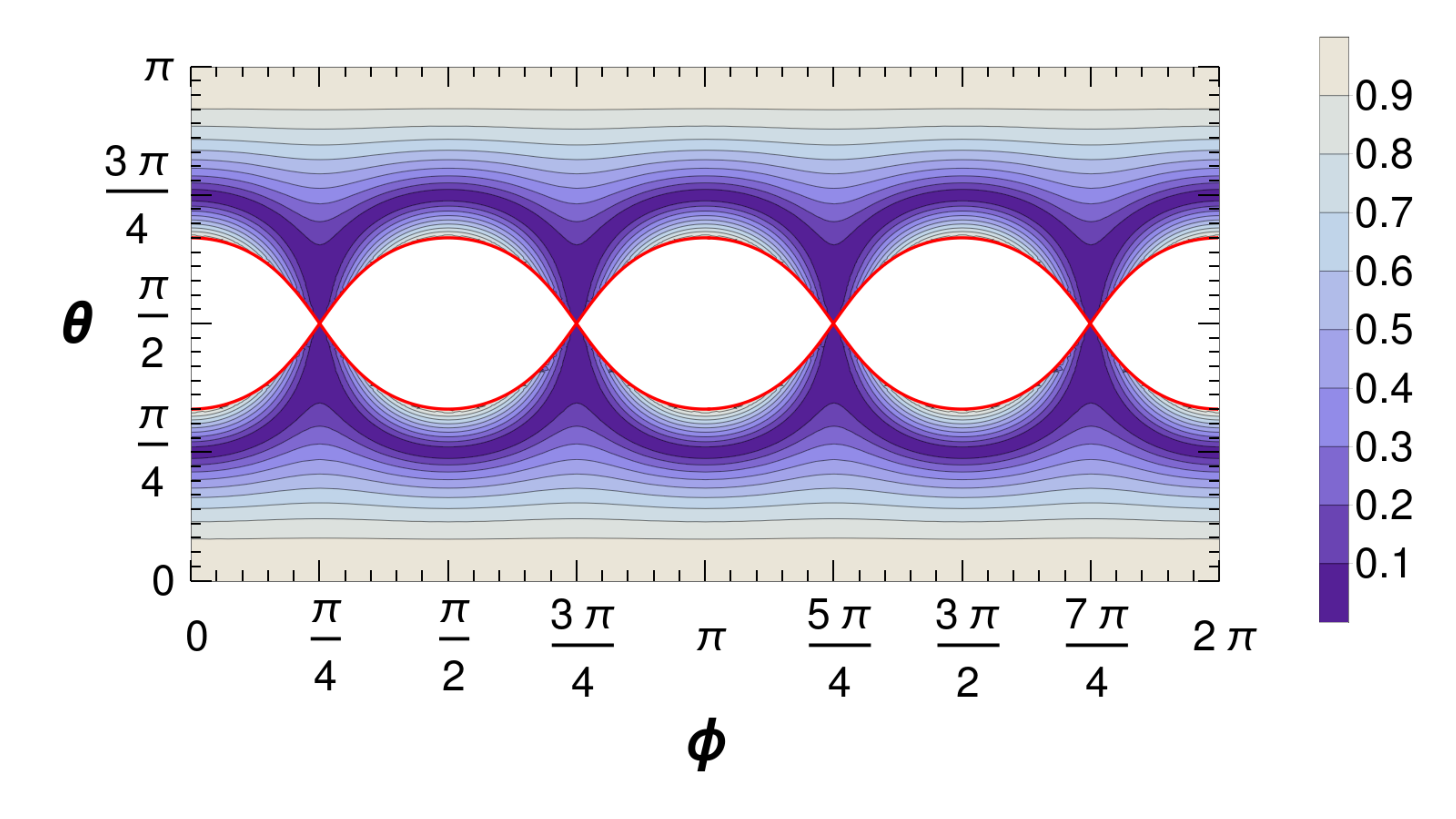}
    \caption{ (Colour online) Volume of the parallelpiped defined by spins 1, 2, and 3.  Planar states are those indicated by the minimum volume (darkest) regions. }
    \label{fig:gsvol}
\end{figure}

To gain insight into what choices of ${(\theta, \phi)}$ pairs give rise to particular states, consider the contour plot in Fig.~\ref{fig:gsvol}. It illustrates the volume of a parallelepiped formed by the $\vec{\mathbf{S}}_1$, $\vec{\mathbf{S}}_2$, and $\vec{\mathbf{S}}_3$ spin vectors corresponding to a particular ${(\theta, \, \phi)}$ chosen from Fig.~\ref{fig:degenplanefull}. It is from this illustration that one can obtain an understanding of what choices of ${(\theta, \, \phi)}$ result in planar states or non-planar states.  Planar states are those having zero volume of the parallelepiped which appear as the darkest regions in  Fig.~\ref{fig:gsvol} as nearly circular rings with a radius of roughly $\pi/4$.  
Note that the nodes indicated on Fig.~\ref{fig:degenplanefull} and Fig.~\ref{fig:gsvol} represent planar states.   
 In terms of the previous polar coordinates planar states can be shown to satisfy trigonometric relations, and, e.g., for $0 \le \phi \le \pi/4$  
the locus of planar states is given by,
\begin{eqnarray}\label{eq:planar}
\theta = \pi - \arcsin{\left(\frac{1}{\sqrt{2} \cos{\phi}}\right)}, 
%\phi &=& \frac{\pi}{4} \cos \gamma \\
%\theta &=& \frac{\pi}{4} \sin \gamma + \frac{n\pi}{2},
\end{eqnarray}
\noindent
%where $n$ is an integer and $\gamma$ varies
outlining a section of the darkest rings in Fig.~\ref{fig:gsvol}. 

\begin{figure}[H] 
    \centering
    \includegraphics[width=1\columnwidth, keepaspectratio]{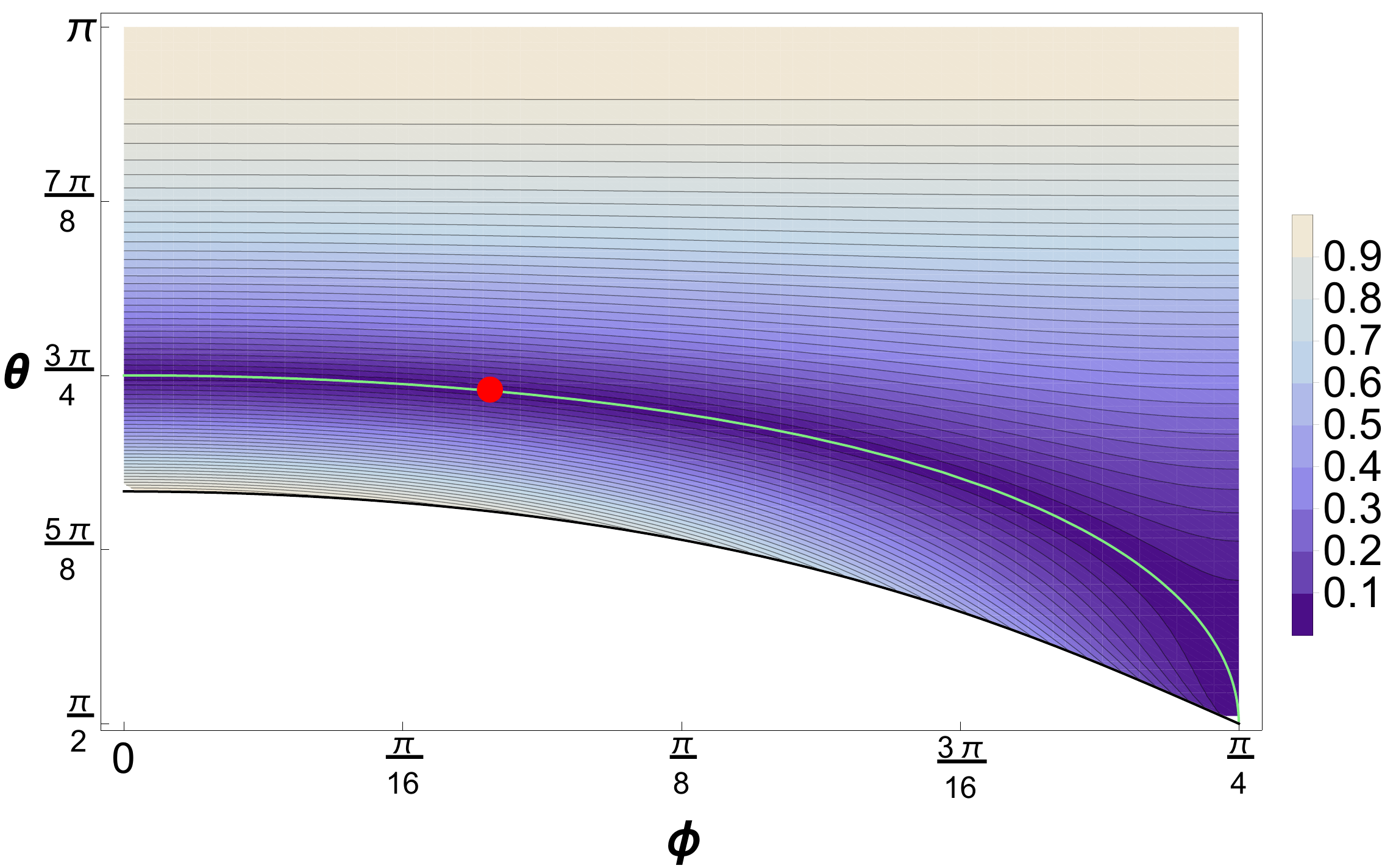}
    \caption{ (Colour online) A segment from Fig.~\ref{fig:gsvol} showing planar state regions (darkest), with the red dot indicating the example shown in Fig.~\ref{fig:PlanarT0b}.   A light curve traces out  Eq.~(\ref{eq:planar}). }
    \label{fig:PlanarT0}
\end{figure}

\begin{figure}[H] 
    \centering
    \includegraphics[width=0.6\columnwidth]{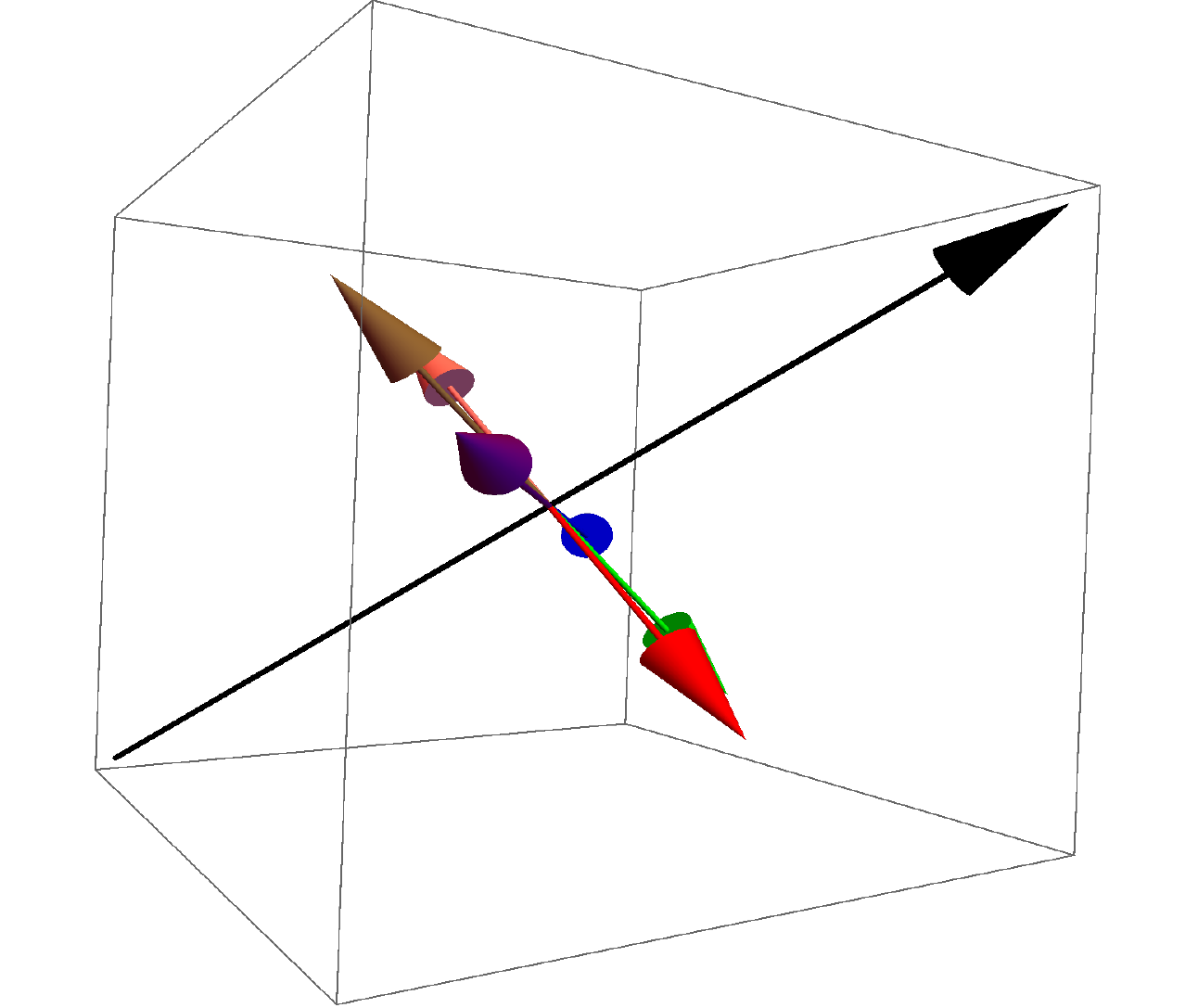}
    \caption{ (Colour online) Example zero temperature planar state (corresponding to the red dot ($\theta=2.324, \phi = 0.258$) shown in Fig.~\ref{fig:PlanarT0}). The [111] axis is indicated by the long black vector.}
    \label{fig:PlanarT0b}
\end{figure}

%================================================================================================================================================================================

\section{Order-by-Disorder at Low Temperature}\label{sec:finitetemp}

\begin{figure}[H]
    \centering
    \includegraphics[width=0.6\columnwidth, keepaspectratio]{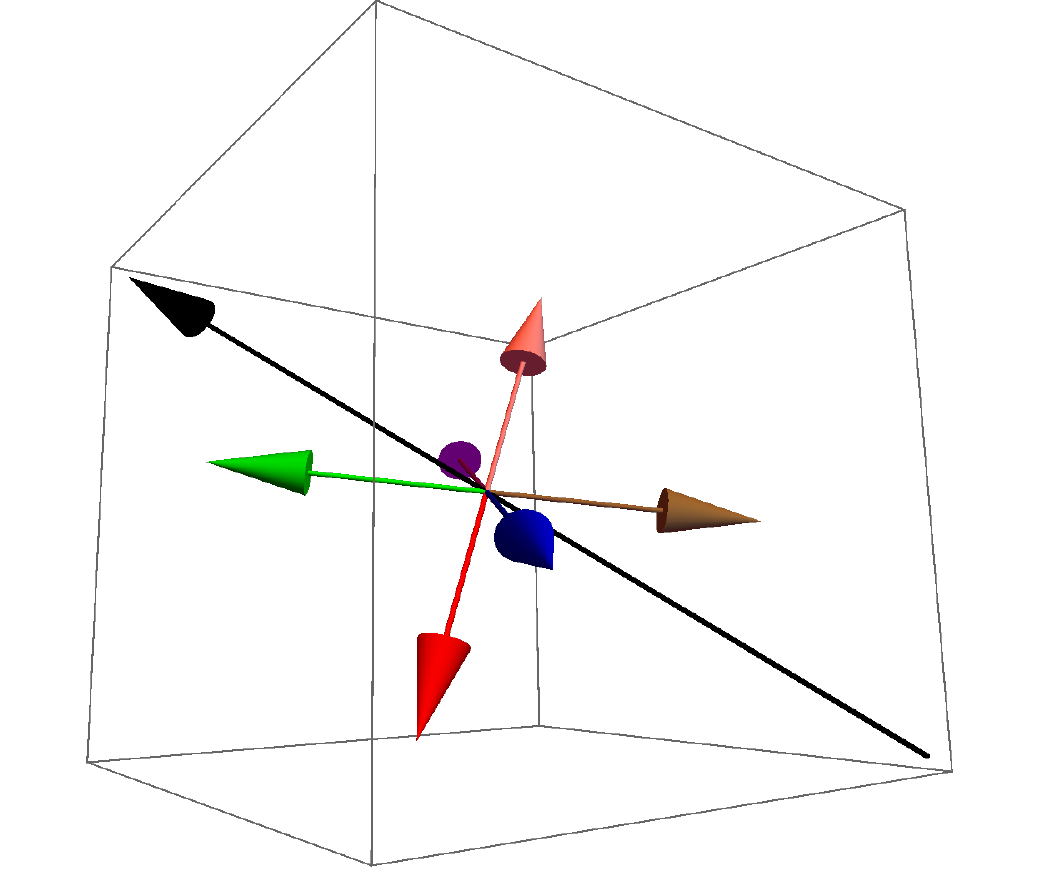}
    \caption{Snapshot of an initial state used in low temperature Heat Bath simulations with $\theta = 2.736, \phi = 0.321$.  The [111] axis is indicated by the long black vector.}% \phi 1.892
    \label{fig:spinsbeforet}
\end{figure}

\begin{figure}[H]
   % \centering
    \includegraphics[width=\columnwidth]{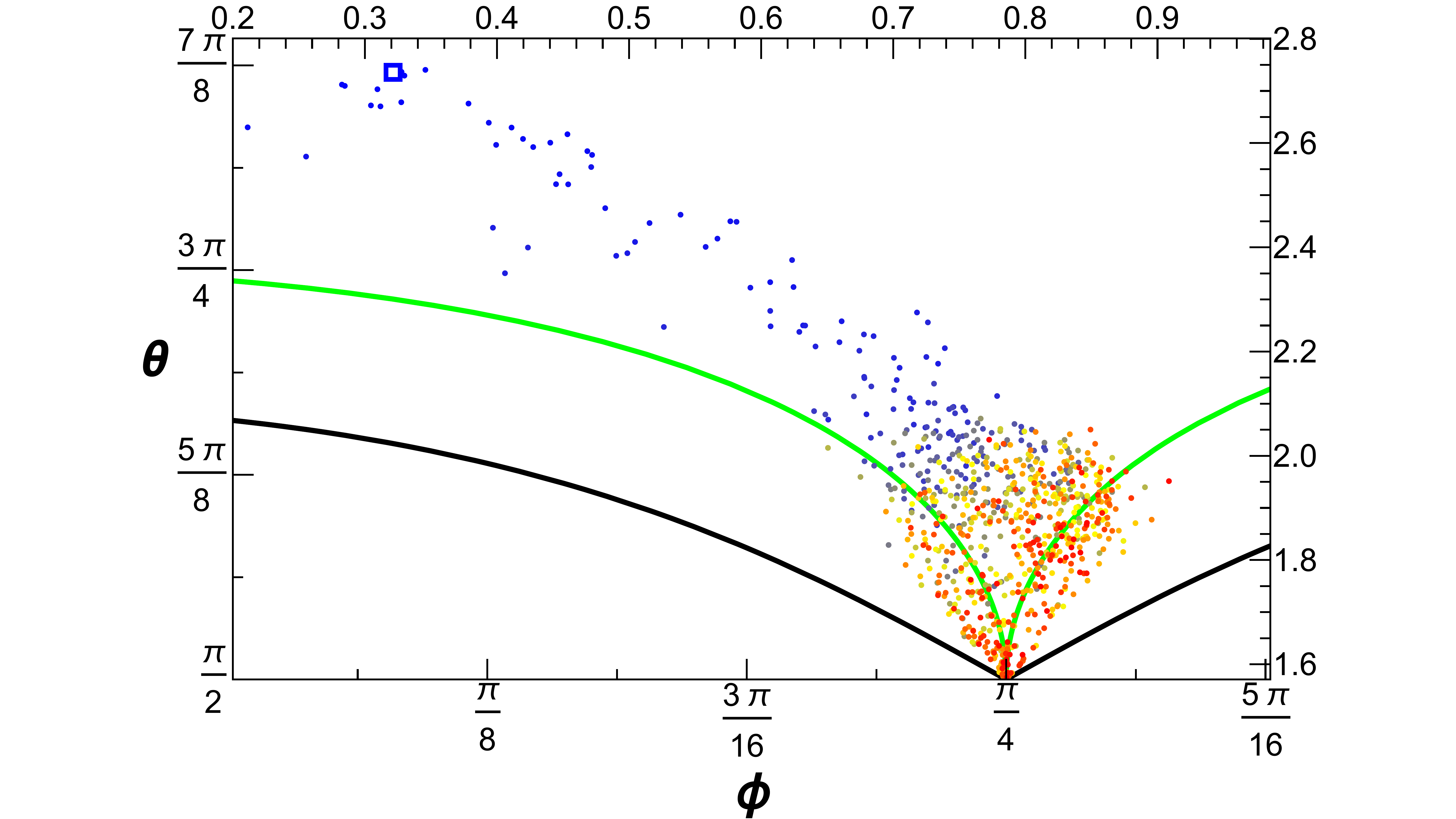}
    \caption{The ground state $\theta$-$\phi$ plane that depicts the evolution states obtained as temperature increased from 0 (blue points) to 0.008 (red points) in increments of 0.00001 from Heat Bath simulations (followed by T=0 EFM simulations) on a lattice with $L = 12$. Black curves mark boundaries for allowed ground states (see Eq.~\ref{eqn:bound}) and green curves are loci of planar states (see Eq.~\ref{eq:planar}).}
    \label{fig:beforet}
\end{figure}

In this section results are presented on the evolution of ground states under the influence of thermal fluctuations.  A low temperature Monte Carlo Heat Bath method,\cite{heatbath} that uses the local field determined through the EFM algorithm, was employed for this purpose.  Beginning with the particular ground state shown in Fig.~\ref{fig:spinsbeforet}, the temperature was increased from $T=0$ to $T=0.008$ with increments of $\Delta T=0.00001$. After each increase in temperature, the spin configurations were run through the $T=0$ EFM to obtain the nearest corresponding ground state. Each of the states obtained following each zero temperature EFM run were recorded and are displayed in Fig.~\ref{fig:beforet}.

\begin{figure}[H]
    \centering
    \includegraphics[width=0.6\columnwidth, keepaspectratio]{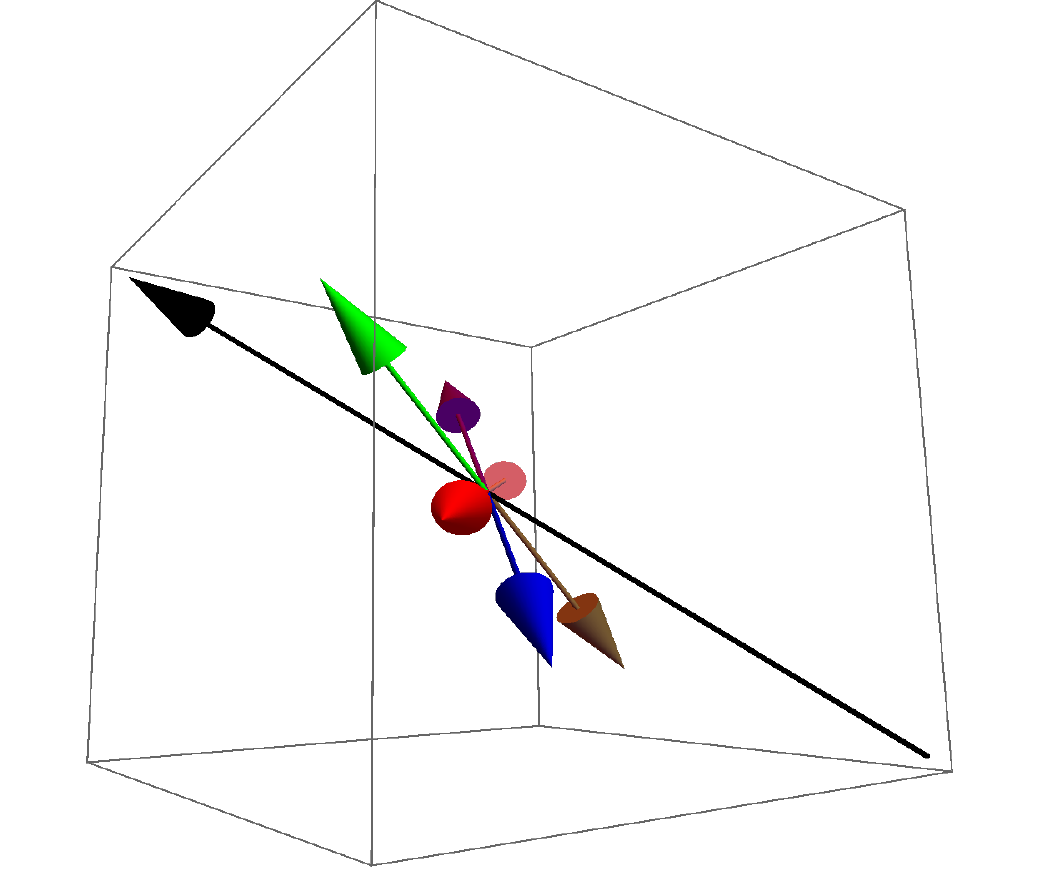}
    \caption{Snapshot of an final state at $T = 0.008$ used in low temperature Heat Bath simulations (Fig.~\ref{fig:beforet}) with $\theta = 1.799, \phi = 2.397$.  The [111] axis is indicated by the long black vector.}
    \label{fig:spinsafter}
\end{figure}

The initial state is clearly a non-planar state. As temperature increases, the spin configurations gradually transition from the non-planar region to the planar region in Fig.~\ref{fig:beforet}, illustrated also by the snapshot of the state at $T = 0.008$ shown in Fig.~\ref{fig:spinsafter}. This order-by-disorder process appears to select the planar states \cite{obd}, near the ``node'' of the bounded region, as the minimum free energy configuration from all possible states after exposing the spin lattice to thermal fluctuations. 

%\begin{figure}[H]
%    \centering
%    \includegraphics[width=0.43\textwidth, keepaspectratio]{img_Andrew/finalt.png}
%    \caption{The ground state $\theata - \phi$ plane that depicts the different states obtained as temperature was varied. Each green dot corresponds to a state that was subject to a heat bath and then the EFM. The state obtained at T=0.00289 is highlighted in red.}
%    \label{fig:aftert}
%\end{figure}

Increasing the temperature results in the spin configuration moving towards a planar node state regardless of its initial state. Temperature has the effect of lifting the degeneracy of the system from one that includes all possible (valid) points in the ground-state manifold to those that are planar (or near planar).

%=================================================================================================================================================================

\section{Applied Magnetic Field}\label{sec:appliedfield}

In this section the effect of an applied magnetic field at zero temperature on the spin configurations is presented.
We are motivated to do so for two reasons: testing whether the field lifts the degeneracy of ground state, and determining structural or phase changes en route to eventual saturation.
Although thermal fluctuations lift the degeneracy, cooling from higher $T$ can result in quenched-in states comprising domains of different ground states, as we show in the next section, and so an external field may provide a way of annealing the system. 
Our first set of results stem from seven EFM simulations performed 
on lattices of size $L=12$, all starting with the same non-planar ground state characterized by $\theta=0.2067$ and  $\phi=3.116$, with the Zeeman term,
\begin{equation}\label{eq:ZeemanHamiltonian}
    E_Z =  - \vec{H} \cdot \sum_i \vec{S}_{i},
\end{equation}
added to the dipole energy, Eq. (1). The magnetic field was increased in steps of $\Delta H =0.0001$ up to $H=0.0150$ and then larger steps of $\Delta H=0.004$  until the field
reached $H\ge0.20$  % $H=0.2150$ 
where the spins begin to approach saturation. Subsequently, the magnetic field
for each lattice was decreased in steps of $H=0.001$ until the 
magnitude was zero (field cycling). Following every change in magnetic field, the spins
  were subjected to the EFM method using 3000 iterations. The magnetic field
directions used for the simulations were along directions of high crystal symmetry;
[001], [010], [100], [011], [101], [110], and [111]. 

\begin{figure}[H]
\centering
\includegraphics[width=\columnwidth]{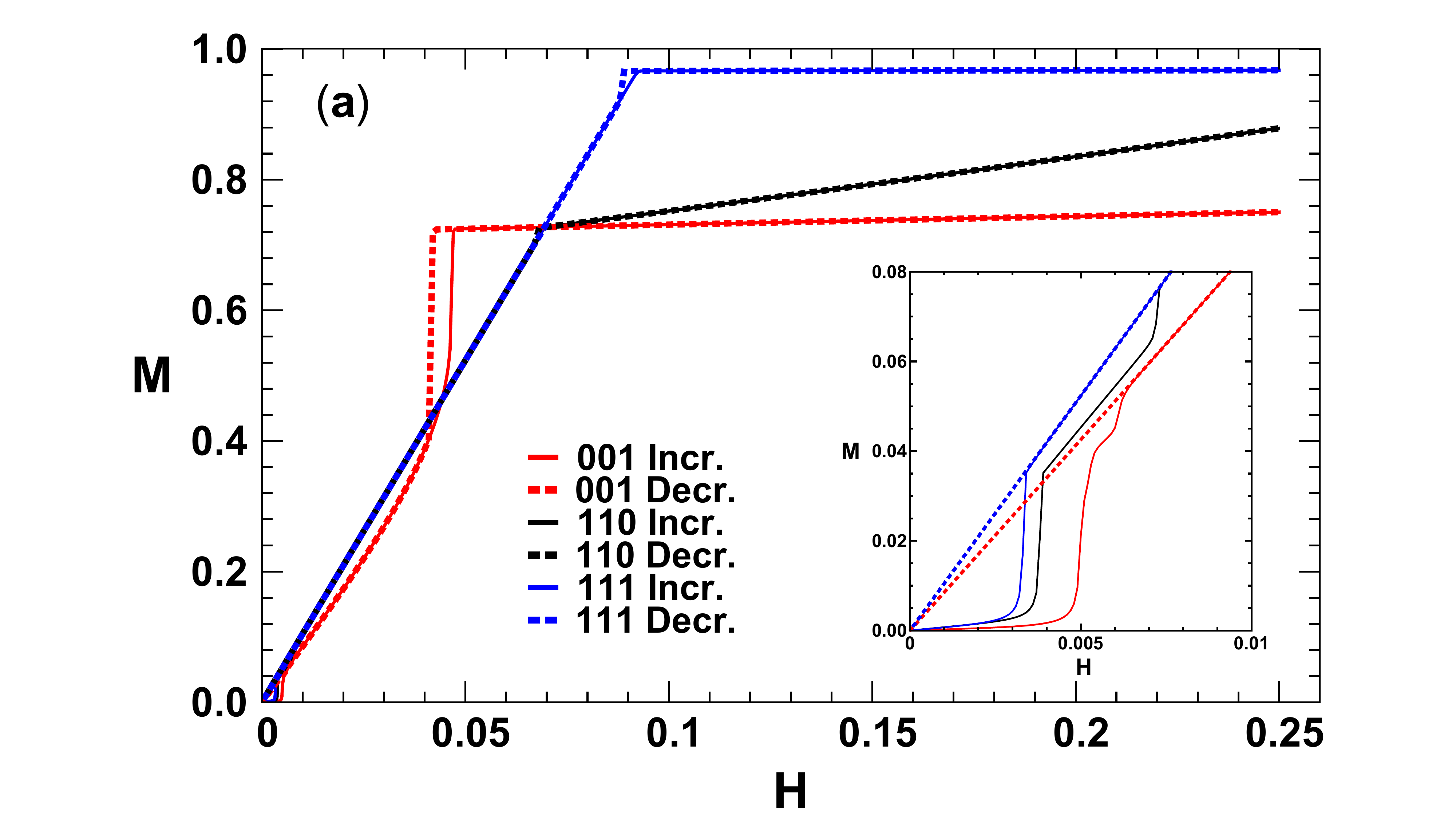}
\includegraphics[width=\columnwidth]{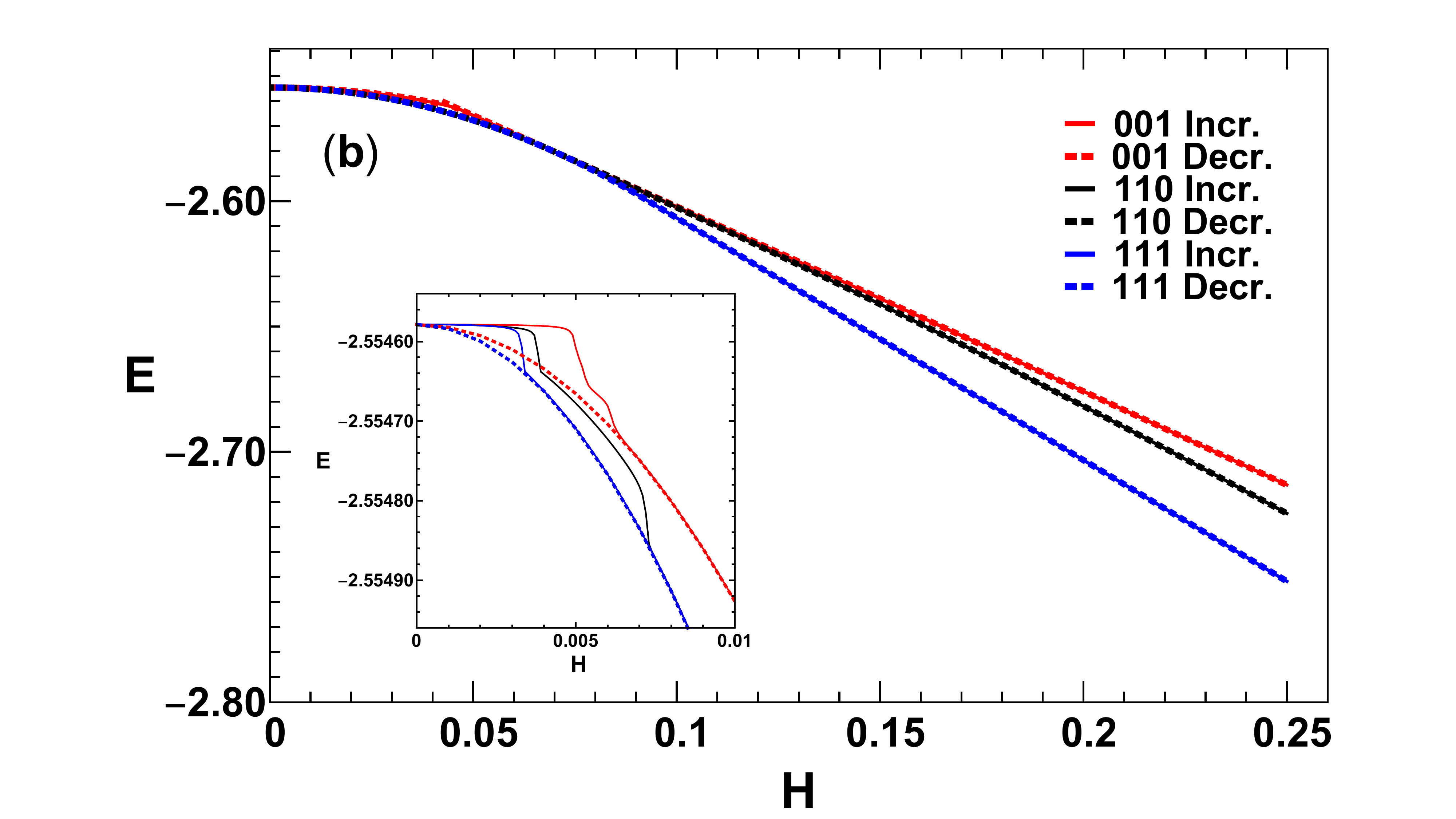}
\includegraphics[width=\columnwidth]{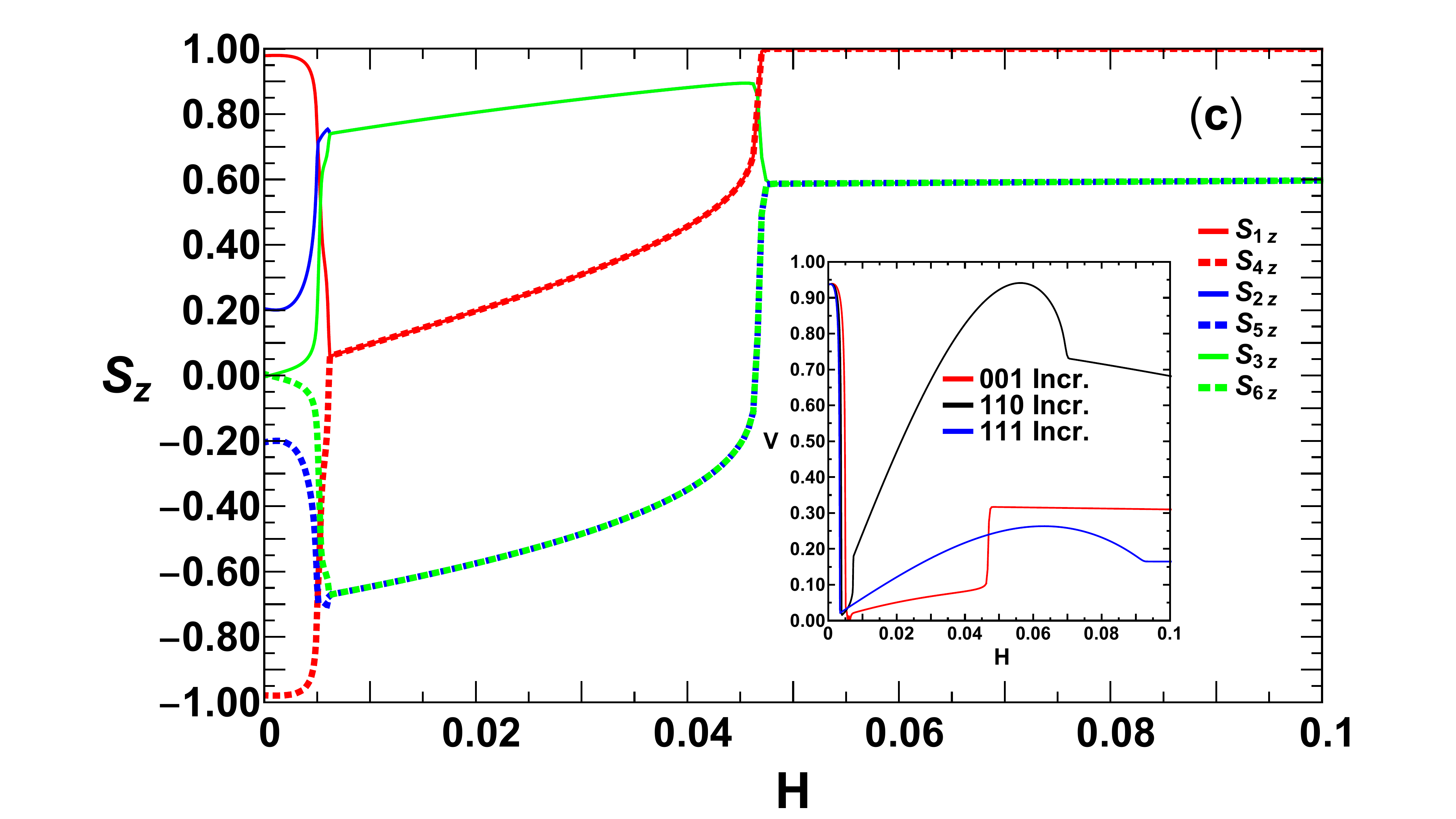}
\caption{Response of a non-planar ground state to external field.  (a) Magnetization for both decreasing and increasing field magnitude, $\vec{\mathbf{H}}\parallel$~[001], [110], and [111].  Hysteresis at higher field is observed only in the [001] and [111] cases. Inset highlights the low-field sudden change in the magnetization, where hysteresis is present for all field directions. (b) The energy as a function of field, with inset showing hysteresis at small field values.  Marked changes in magnetization are accompanied by changes in the slope of the energy with $H$. (c) $z$-components of the unit spin vectors for the six sublattice spins when $H$ is increased along the [001] direction, showing the pairing of sublattice spins at transitions.  Inset shows the volume of the parallelepiped formed by $\vec{S}_1$, $\vec{S}_2$, and $\vec{S}_3$, which is near zero for a canted planar state.} 
\label{fig:001compound}
\end{figure}

\begin{figure}%[H]
    \centering
     \begin{subfigure}[t]{0.5\columnwidth}
        \centering
        \includegraphics[width=0.75\columnwidth]{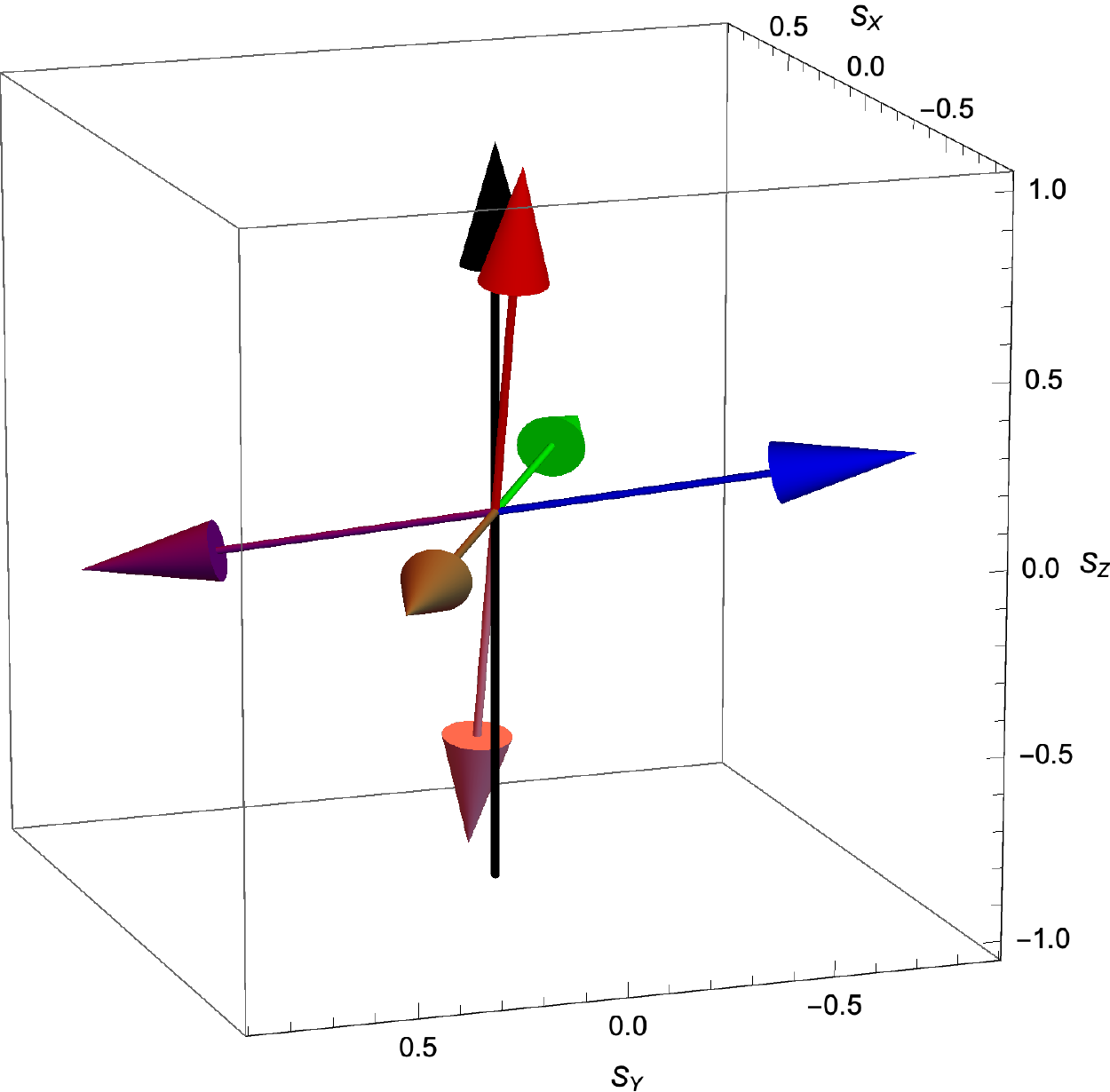}
        \caption{H=0}
    \end{subfigure}%
    ~ 
     \begin{subfigure}[t]{0.5\columnwidth}
        \centering
        \includegraphics[width=0.75\columnwidth]{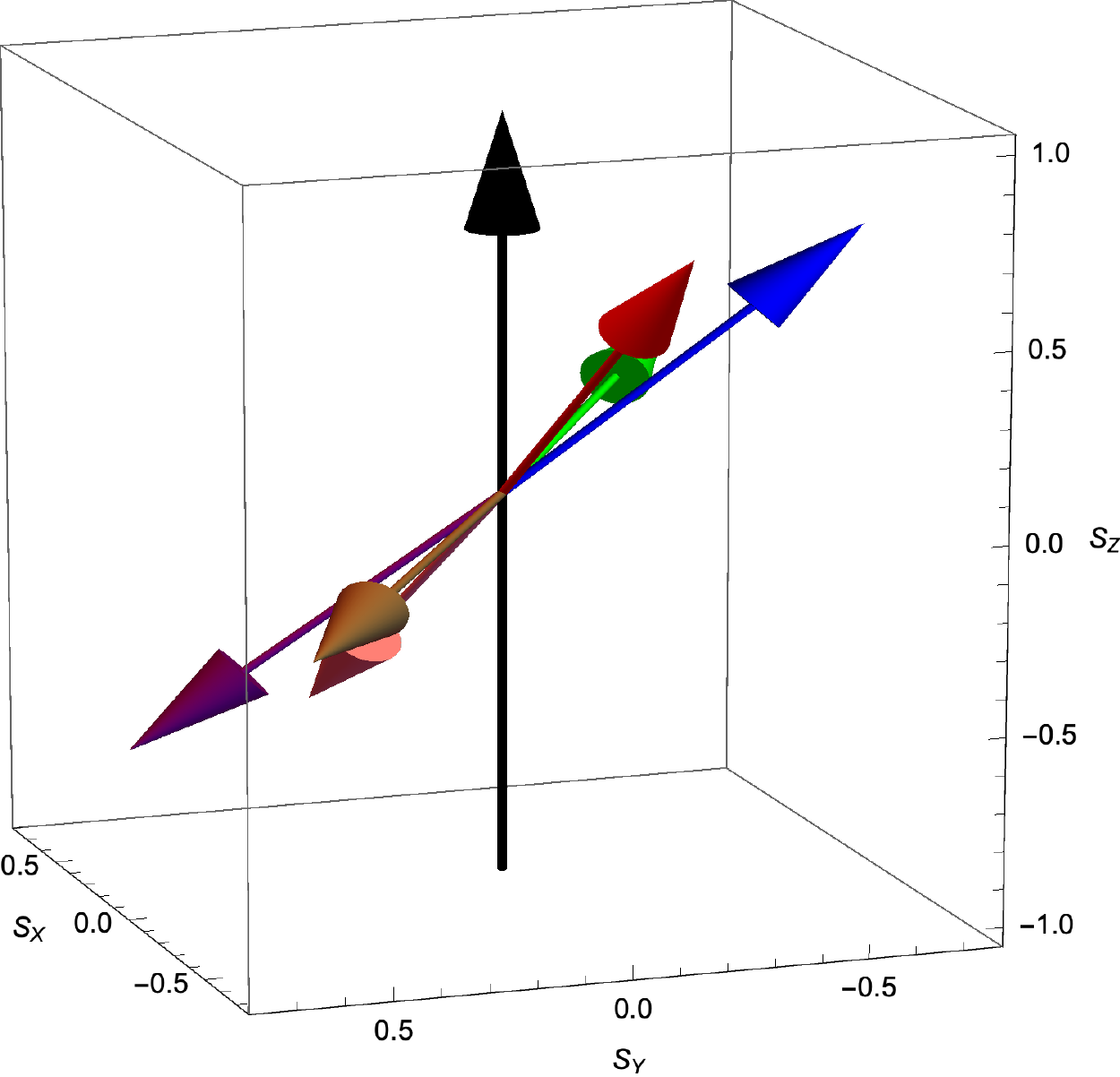}
        \caption{H=0.0050}
    \end{subfigure}
    \begin{subfigure}[t]{0.5\columnwidth}
        \centering
        \includegraphics[width=0.75\columnwidth]{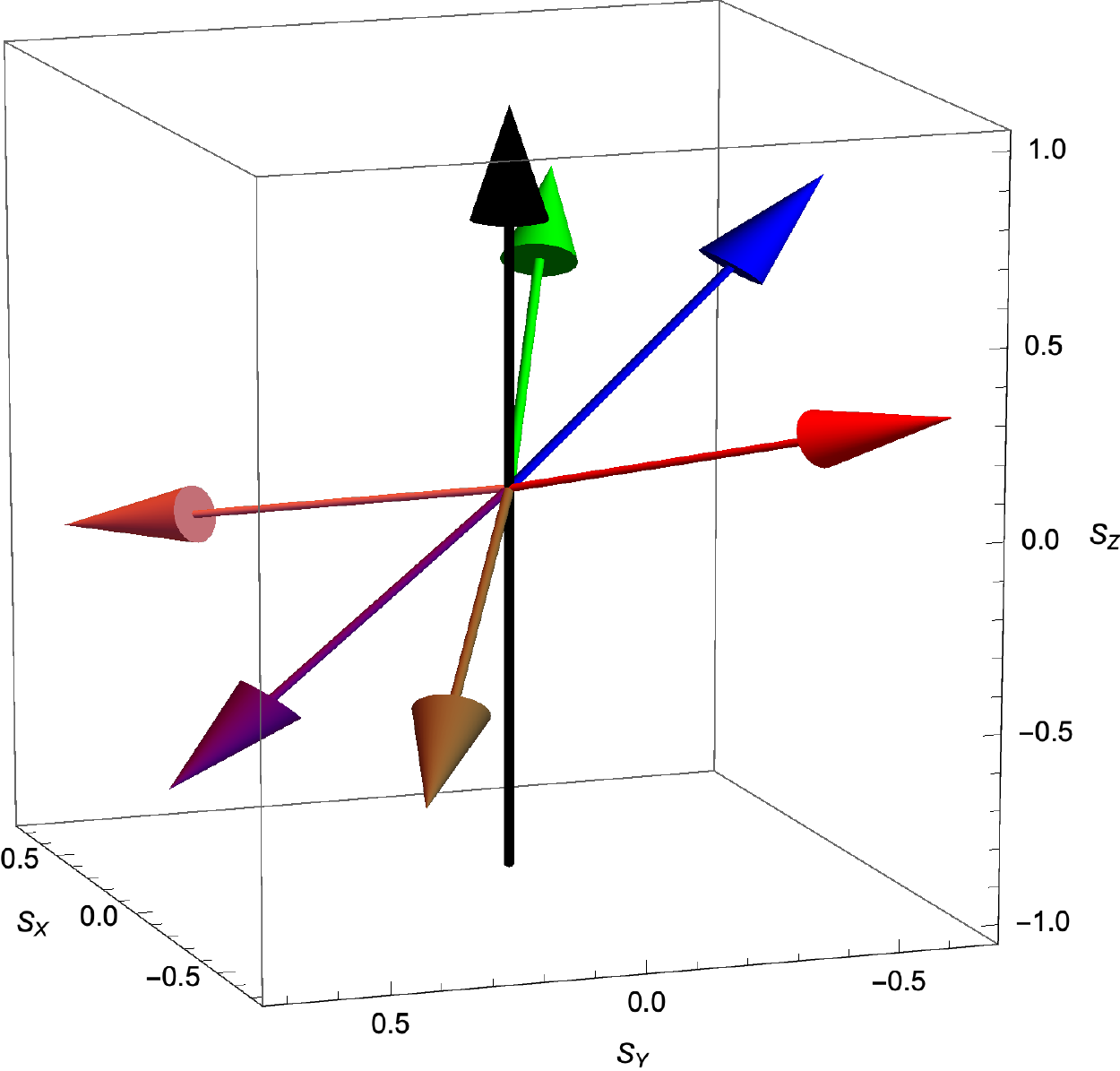}
        \caption{H=0.0058}
    \end{subfigure}%
    ~ 
     \begin{subfigure}[t]{0.5\columnwidth}
        \centering
        \includegraphics[width=0.75\columnwidth]{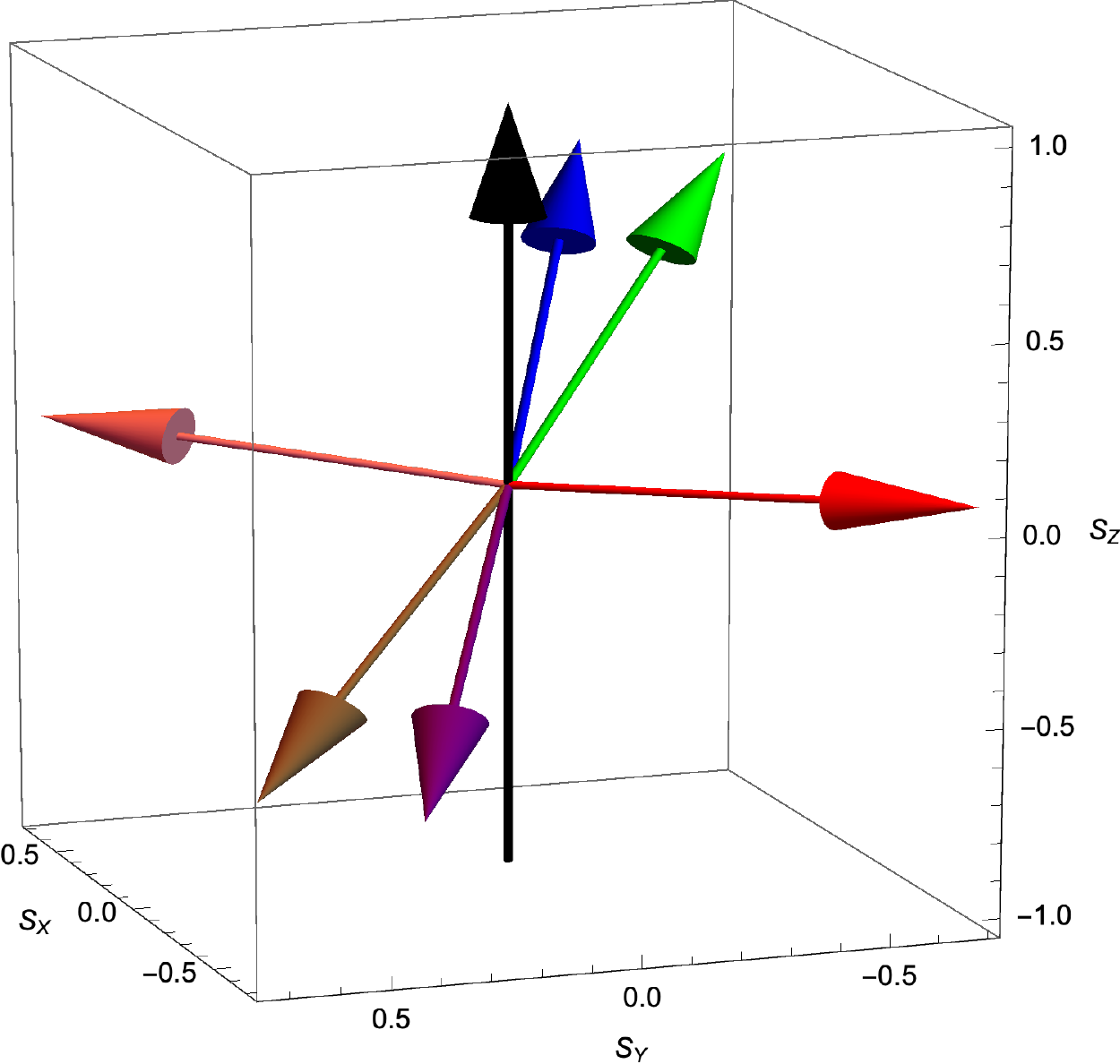}
        \caption{H=0.0073}
    \end{subfigure}
    \begin{subfigure}[t]{0.5\columnwidth}
        \centering
        \includegraphics[width=0.75\columnwidth]{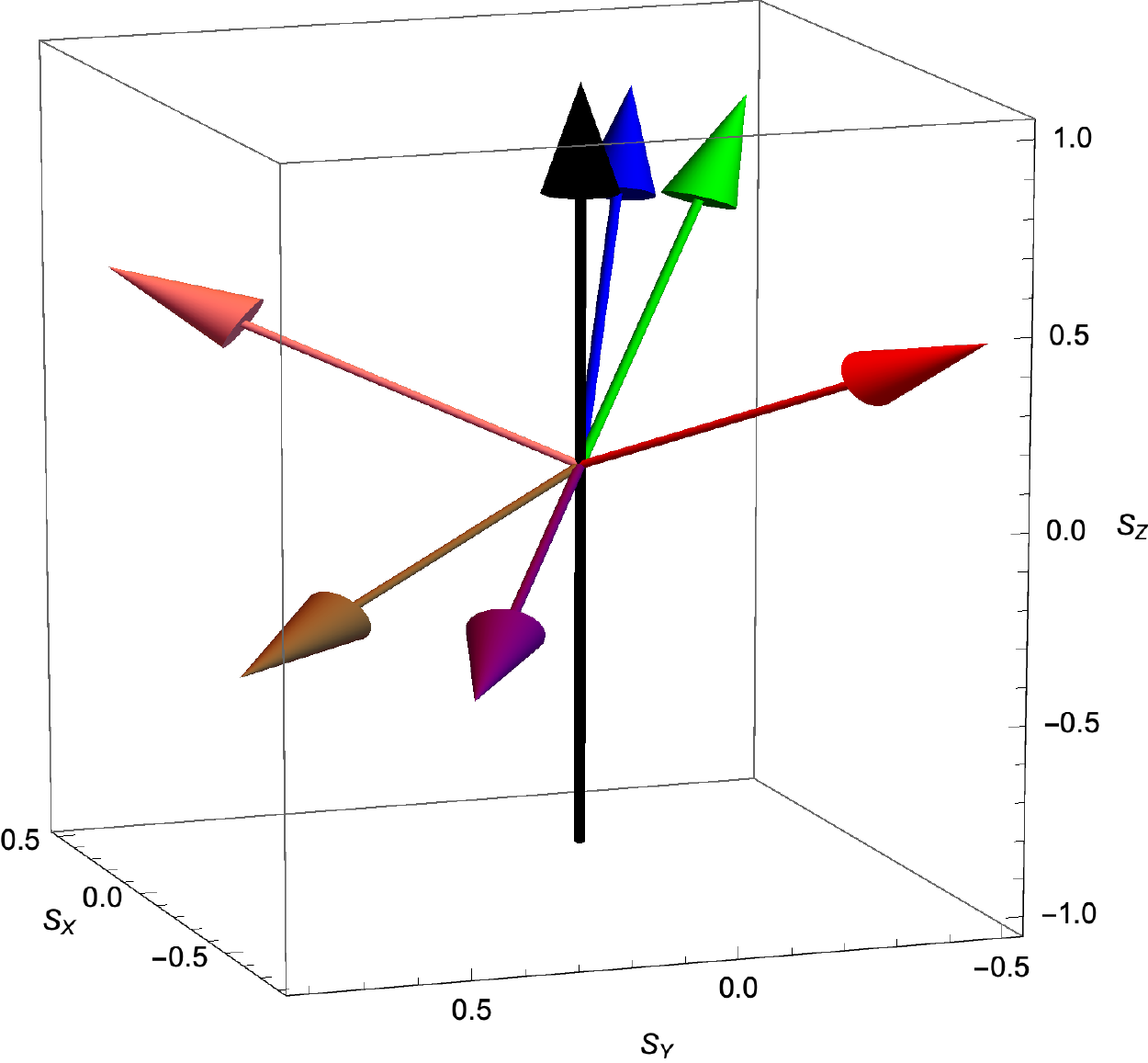}
        \caption{H=0.0402}
    \end{subfigure}%
    ~ 
     \begin{subfigure}[t]{0.5\columnwidth}
        \centering
     \includegraphics[width=0.75\columnwidth]{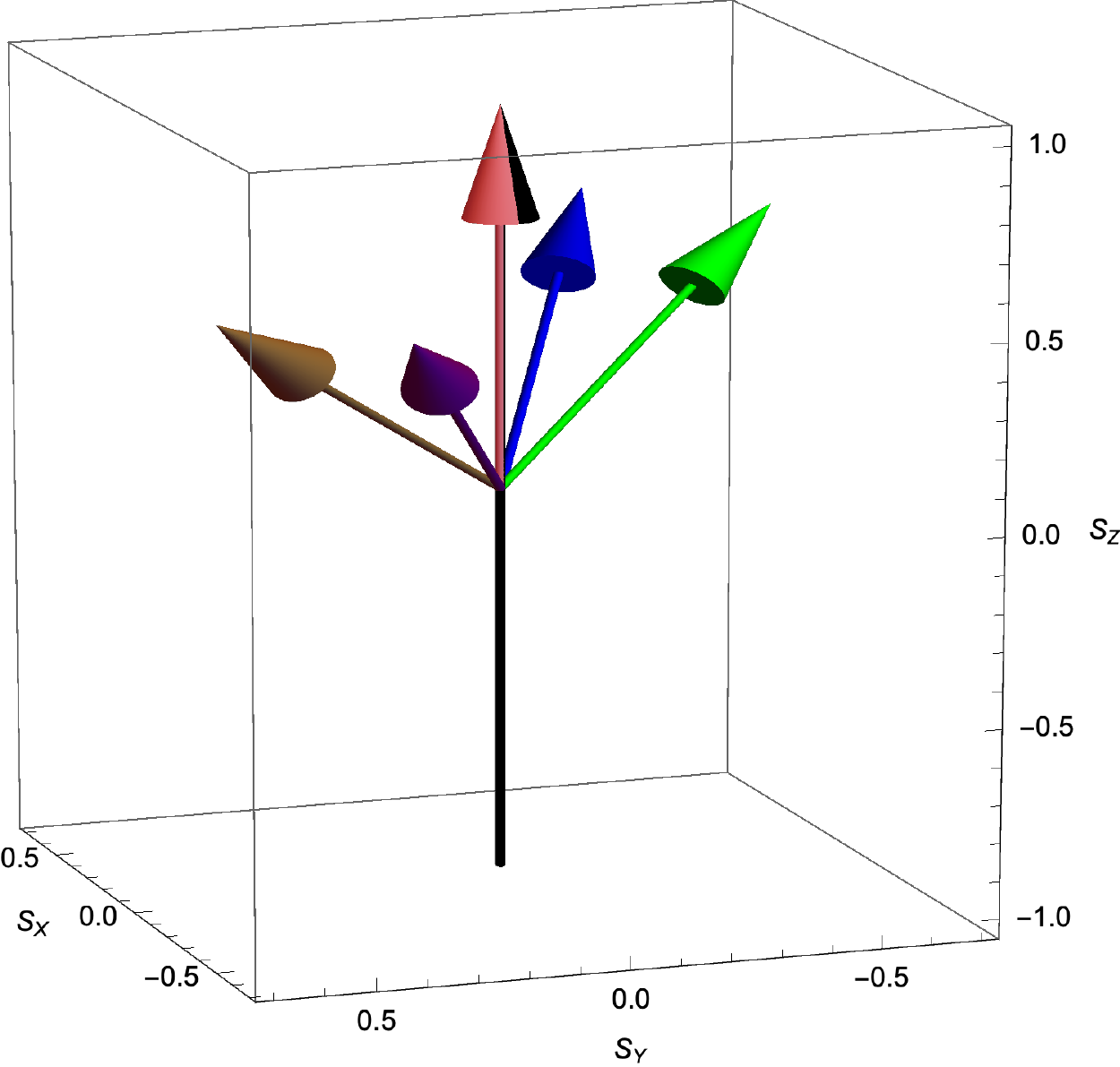}
        \caption{H=0.0602}
    \end{subfigure}%
    \caption{Snap shots of the six sublattice spins  obtained on increasing the external field from H=0 to H=0.0602. The black arrow indicates the direction of the field along [001].}
  \label{fig:001vecsinc}
\end{figure}

Illustrative results for the initial non-planar ground state above with the field along principal cube axes are shown in Fig.~\ref{fig:001compound}, where
the magnetization, energy and spin components as functions of magnetic field indicate several distinct phases.  Snapshots of the six sublattice spin vector orientations with increasing field are shown in Fig.~\ref{fig:001vecsinc}.

The initial non-planar ground state has zero magnetization and Fig.~\ref{fig:001compound}(a)-inset shows that  the magnetization lingers close to zero at low values of the magnetic field along the three directions. 
At some critical value (H $\approx$ 0.005) with $\vec{\mathbf{H}}\parallel$~[001], sudden changes in both the magnetization and energy occur [insets of Figs.~\ref{fig:001compound}(a) and (b)] as well as in the six sublattice magnetizations [Fig.~\ref{fig:001compound}(c)], signaling the formation of a `planar' state that is canted by the field.
The sublattice spin configurations before and after this change are shown in Fig.~\ref{fig:001compound}(c) and Fig.~\ref{fig:001vecsinc}(a) and (b). 
%for all the non-planar states 
%{\color{red} [Isn't there only one state that these fields are applied to?]}
%after which a planar state was formed.  
Upon increasing the field further, the plane of spins rotates [Fig.~\ref{fig:001vecsinc}(c)] until there is a sudden switch between two of the sublattices at $H \approx 0.0073$ [Fig.~\ref{fig:001vecsinc}(d)], where again the magnetization and energy change suddenly. % [Fig.~\ref{fig:001compound}(a)-inset and Fig.~\ref{fig:001compound}(b)-inset]. 
This (canted) planar state is characterized by the equality of the $z$-components of pairs of spins $\vec{S}_1$ and $\vec{S}_4$, $\vec{S}_2$ and $\vec{S}_3$, and $\vec{S}_5$ and $\vec{S}_6$, as shown in Fig.~\ref{fig:001compound}(c), and a small volume of the parallelepiped formed by $\vec{S}_1$, $\vec{S}_2$, and $\vec{S}_3$, as shown in the inset of Fig.~\ref{fig:001compound}(c).
A similar jump to a planar state occurs for all the field directions, 
although for the field along [111] the transition happens in one step rather than in two. The formation of a planar state is reminiscent of the spin-flop transition in antiferromagnetic systems with anisotropy, for which application of a field results in spins flopping down into the plane perpendicular to the easy axis and then canting in the field direction. \cite{Bogdanov:2007he}

Beyond $H \approx 0.0073$, the magnetization increases in an approximately linear fashion as the canting of the spins increases.  At $H \approx 0.045$, there is another sudden change in magnetization as sublattices $\vec{S}_1$ and $\vec{S}_4$ align with field, and the remaining sublattices lock into having the same $z$ component.  The magnetization of this final state, with two sublattices aligned with the field and four canted, increases slowly with increasing field strength.  (Note there is a slight positive slope in the plateau regime.)  Similar pairing up of sublattices occurs for $H$ along other directions, although the details differ.  For example, for $H$ along [110], no pair in the final state aligns directly with field, and for  $H$ along [111], in the final state all sublattices share the same component along the field.

%\begin{figure}[H]
%\centering
%\includegraphics[width=0.43\textwidth, keepaspectratio]{img_Andrew/magneticphases.png}
%\caption{Example of magnetization with a magnetic field oriented along the y-axis.}
%\label{fig:magphase}
%\end{figure}

%Fig.~\ref{fig:001compound} illustrates the magnetization differences between increasing and decreasing the magnetic field. The differences are subtle, but highlight the existence of hysteresis in this system. 

%\begin{figure}[ht!]

For the field in the [001] direction, there is a pronounced hysteresis  near the transition to the saturation plateau ($H \approx 0.045$), which is more weakly present for $H$ along [111]  and largely absent for $H$ along [110].  These MH loops can also be dependent upon which of the crystallographically equivalent high symmetry directions are chosen. 
%For example, three different curves can occur [how are they different?  some have loops, some not?] with H increasing along [100], [010], and [001], dependent upon which of the seven ground states was chosen as a starting configuration. [seven starting ground states?]}  
For example, three different MH curves can occur for a given starting ground state with H increasing along [100], [010], and [001].
However, all of the crystallographically equivalent directions show identical MH curves upon decreasing the field from saturation. These behaviours indicate that the energetically equivalent ground states experience a complex energy landscape as a function of applied field.

These simulations reveal how a sufficiently large external magnetic field causes a spin configuration to transition from a non-planar state to a planar state, characterized by pairs of sublattices that have the same component along the field direction, and then to a state that will saturate and that has two or more pairs of sublattices arranged to have the same component along the field. When decreasing the magnetic field magnitude, the spin vectors orient themselves away from the magnetic field direction and form a planar state at zero magnetic field. The general response of the spins to the magnetic field are consistent with the exception of one point; spin configurations that are already planar at zero field do not experience an abrupt change into a planar state, but rather a linear change similar to non-planar states after transitioning to a planar state.  An examination of the energy hysteresis occurring in the lowest $H$ regime, shown in the inset of Fig.~\ref{fig:001compound}(b), clearly shows that at finite field the planar state has a lower energy than the non-planar state.   This fact indicates that the non-planar states are metastable in this region.  This lifting of energy degeneracy allows for the development of energy barriers (and hence hysteresis) between configurations that at zero field can move continuously from one to another.  Qualitatively similar results are found for other ground states.  One quantitative difference is the field values at which transitions occur.

\begin{figure}[H]   % Should a reduced ground state be introduced here or later?
    \centering
    \includegraphics[width=0.95\columnwidth]{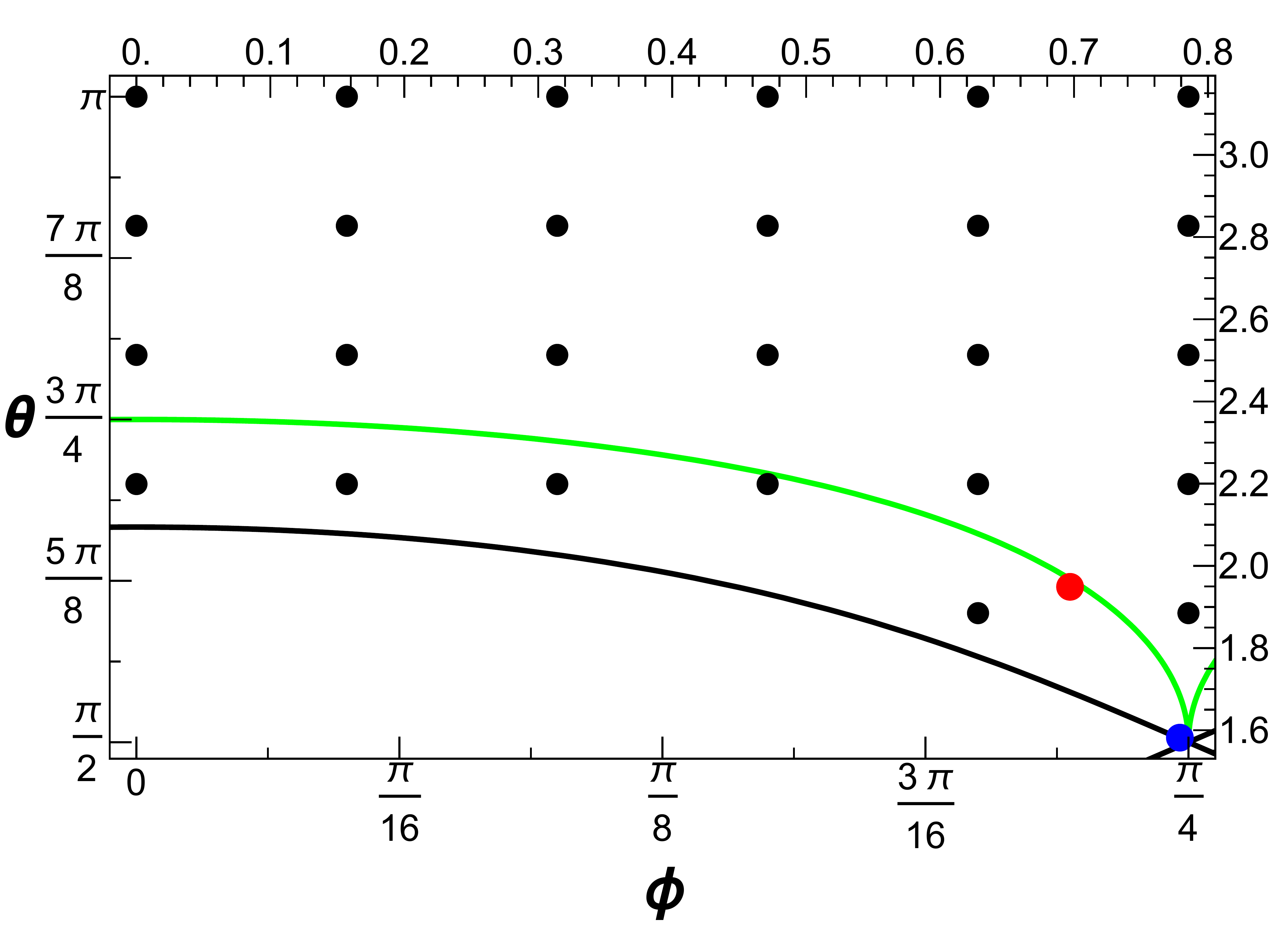}
    \caption{ Black points show the values of the polar angles that identify the twenty-six initial degenerate spin configurations used to study the effect of field, with {\bf H} along the [111] direction.  After field cycling, twenty-four states had the same planar configuration ($\theta = 1.949, \phi = 0.697$), indicated by the red dot. Two other planar states indicated by the blue dot ($\theta = 1.582, \phi = 0.779$) also occurred.  Black and green curves are the same as in Fig.~\ref{fig:beforet}.} 
    \label{fig:groundstatearray}
\end{figure}

%The most notable result of 
Our second set of results show
that all possible ground states can be transformed into planar states through field cycling. Twenty-six different initial ground states spanning $\theta$ and $\phi$ were prepared, as indicated in Fig.~\ref{fig:groundstatearray}. Each of these states was field cycled
%{\color{blue} field-cycled} 
with a field along the [111] direction.

Twenty-four of the initial   non-planar states resulted in the same final state  upon field cycling, %corresponding, 
while only two resulted in a different final state,  a node state, as indicated in Fig.~\ref{fig:groundstatearray}. Both of these are planar states. Thus, it is clear that by cycling an external magnetic field, the degeneracy of the ground state is reduced to include only states that are planar.

%\begin{figure}[H]   % Should a reduced ground state be introduced here or later?
%    \centering
%    \includegraphics[width=0.43\textwidth, keepaspectratio]{img_Andrew/postdegaussedRELABELED.png}
%    \caption{ Points indicate the final two states of the inital twenty-six shown in Fig.~\ref{fig:groundstatearray} after undergoing degaussing with H along [111]. }
%    \label{fig:postdegaussedarray}
%\end{figure}

%\input{montecarlo.tex}

%=======================================================================================================================================================================================

\section{Monte Carlo Simulations}\label{sec:mc}
%    Results from the using Metropolis MC simulation at finite temperatures are described here. Energy and specific heat (C) are among the quantities measured in simulations to determine phase transitions. 
\subsection{Energy and Specific Heat}\label{ssc:energy}

%Paragraph 1
%Examples of the average energies are given in Fig.~\ref{fig:EnergyIT} and Fig.~\ref{fig:EnergyL12} for isolated temperatures at varying lattice sizes and MC cooling and heating simulations, respectively.

As in our previous MC simulations on the 2D kagome lattice,\cite{holden2015} three main types of temperature runs were performed: isolated temperature, cooling and heating.  All simulations were done at zero magnetic field on lattices with $L = $ 6, 12, and 18. Isolated temperature data are collected by a simulation at every $T$ initialized with a random configuration. These initial configurations have no dependence on the configurations of the neighbouring temperatures. Furthermore, each temperature may be run simultaneously on separate processors, greatly decreasing the real time length of simulations. Heating and cooling runs, however, are completed at consecutive $T$ using the final spin configuration of the previous $T$ as the initial configuration.  In the case of isolated temperature runs, at each temperature, at least 9$\times$10$^5$ MC steps (MCS) were used with the initial $1\times 10^5$ MCS discarded for thermalization giving about $N_{\rm MCS}=8\times 10^5$ MCS for averaging.  A single MCS involves, on average, an attempt to flip each spin in the system.   For cooling and heating runs, these numbers were reduced by a factor of ten. Preliminary MC results on the 3D kagome system using a small lattice ($L$ = 8) with a limited number of MCS were previously reported\cite{holden2015,shanethesis} and indicated anomalous behaviour at a temperature near 0.35. 
 
\begin{figure}[H]
        \centering
        \includegraphics[width=1\columnwidth, keepaspectratio]{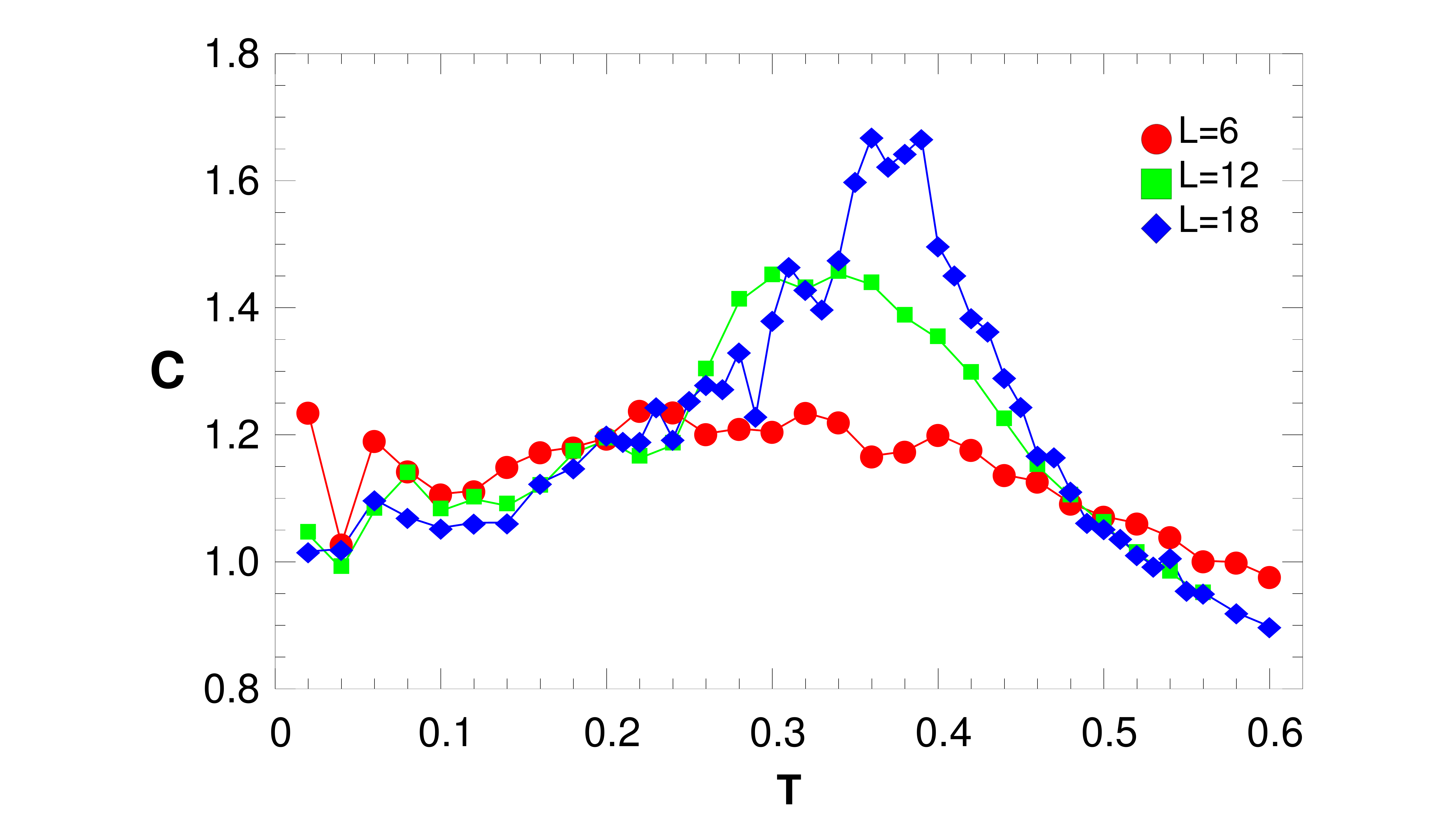}
        \caption{Specific heat for different lattice sizes from isolated temperature MC simulations, averaged over 800,000 MCS at each $T$.}
        \label{fig:HeatIT}
    \end{figure}

Figure~\ref{fig:HeatIT} shows results for the specific heat, given by,
\begin{equation}\label{eq:SpecHeat}
C = \dfrac{1}{N_{\rm MCS}}\sum_{s}{\dfrac{(E_{s} - \left< E \right>)^{2}}{T}}
\end{equation} 
for $L$ = 6, 12, and 18, where the sum is over the MC generated configurations.

In each set there is a peak in the range $T\in[0.20,0.40]$ which becomes larger, narrower, and appears at higher $T$, with increasing system size. The largest and most well defined peak occurs in the $L=18$ case near $T_C=0.38$. At low temperatures, the results suggest $C \simeq 1$, consistent with earlier simulation results.\cite{holden2015}
Note that the classical Heisenberg model behaves anomalously at low temperatures. The model can be solved exactly in one dimension where these is no phase transition \cite{spht} and the specific heat approaches
a constant value as $T\rightarrow 0$. The model behaves like a single spin in an effective magnetic field with two continuous degrees of freedom. The equipartition theorem yields an energy linear in $T$ and a constant specific heat.

\begin{figure}[H]
        \centering
        \includegraphics[width=1\columnwidth, keepaspectratio]{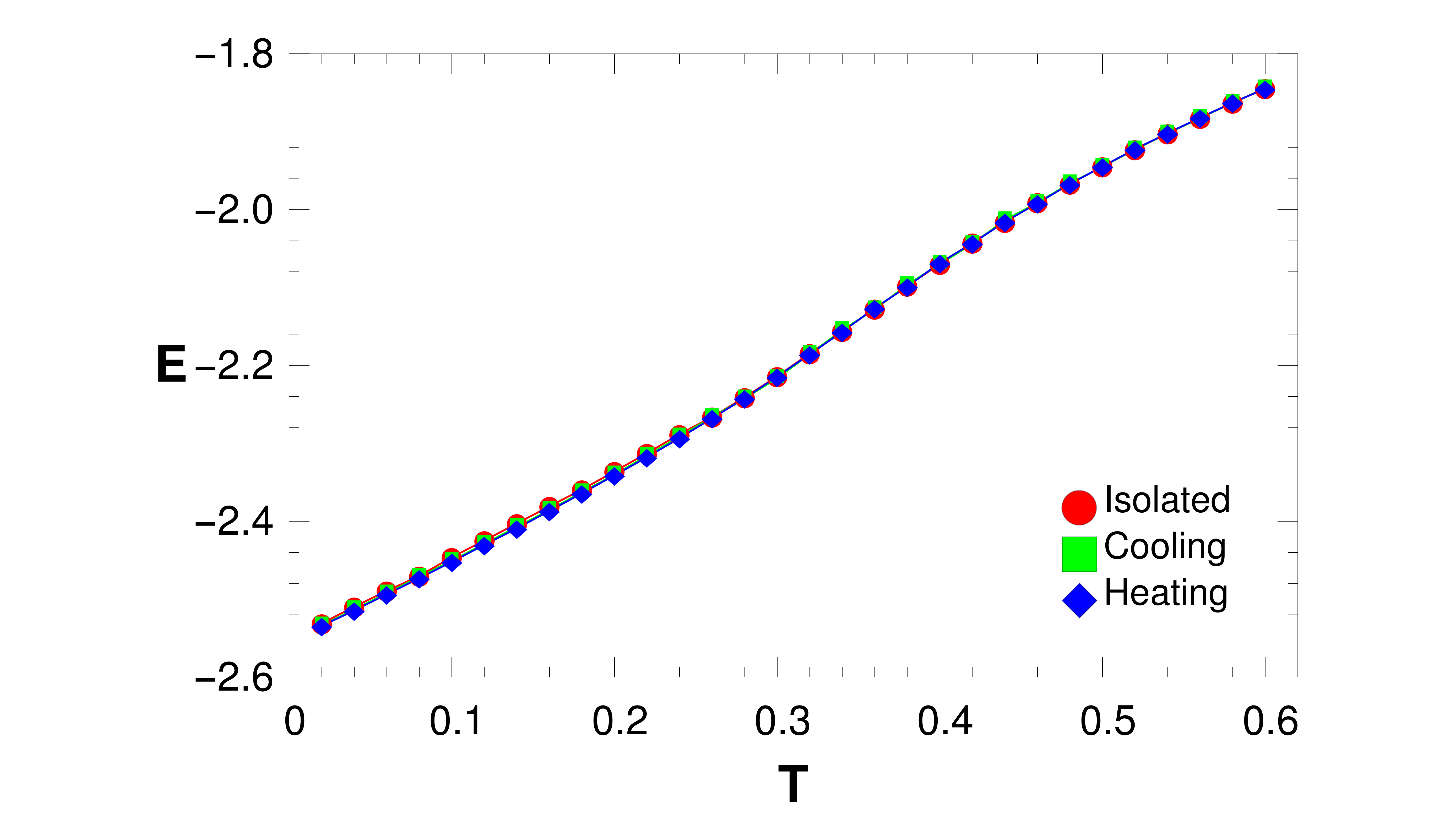}
        \caption[Energy per site as a function of temperature for $L=12$ simulations]{Measured energy per site as a function of temperature for $L=12$ cooling, heating and isolated temperature simulations.  Heating and cooling simulations are averaged over 80,000 MCS at each $T$.}
        \label{fig:EnergyL12}
    \end{figure}

Figure~\ref{fig:EnergyL12} shows the energy at $L$ = 12 performed at isolated temepratures as well as cooling and heating runs.  There appears to be little difference in the simulation types. This observation is important to the results shown below for the sublattice magnetization, which do exhibit a strong dependence on simulation type.    
 
\subsection{Magnetization and Susceptibility}\label{ssc:magnet}

\begin{figure}[H]
        \centering
        \includegraphics[width=1\columnwidth, keepaspectratio]{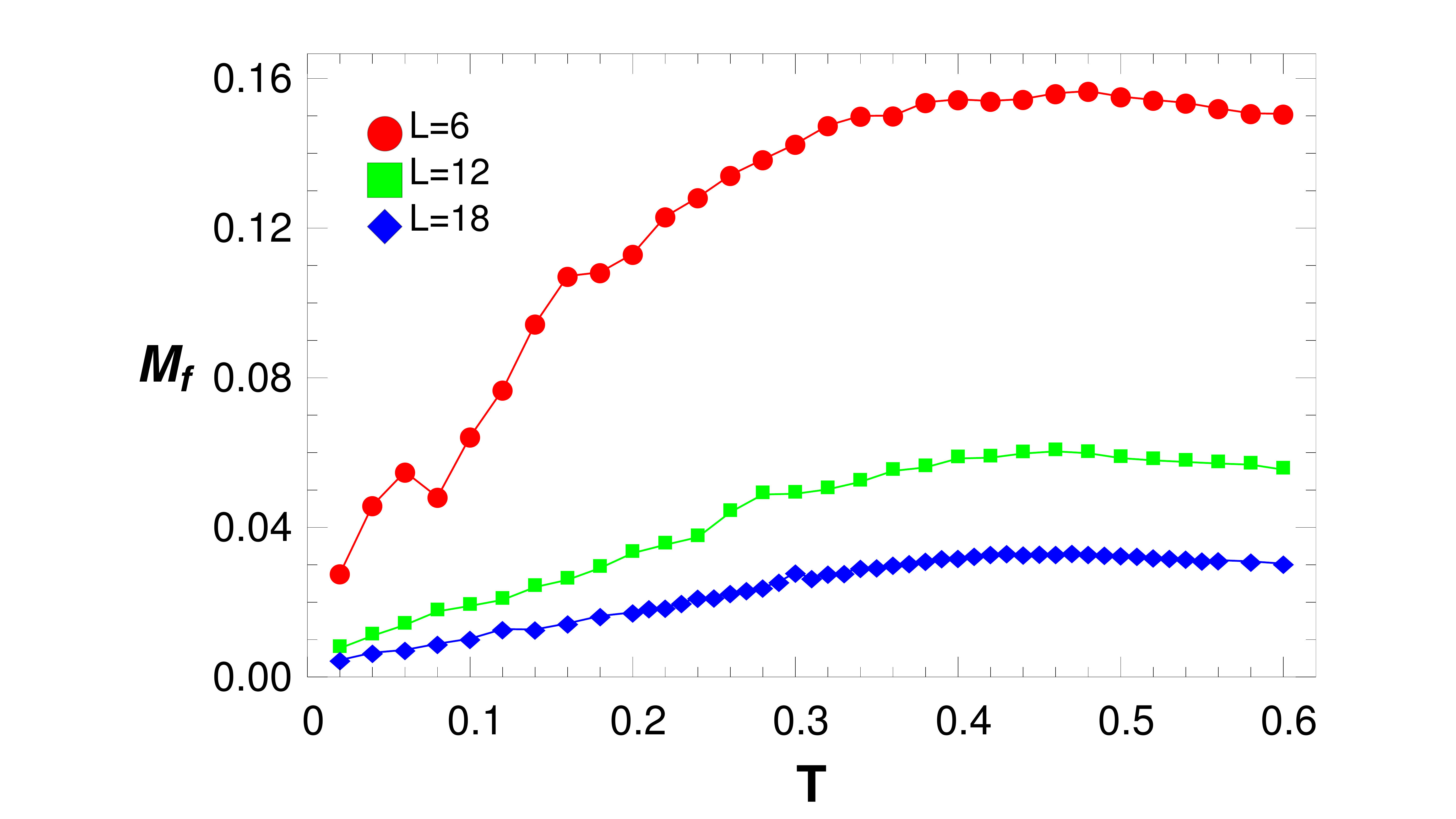}
        \caption[Total ferromagnetic magnetization vs temperature for varying lattice size at isolated temperatures.]{Total ferromagnetic magnetization for different lattice sizes $L$ from isolated temperature MC simulations.}
        \label{fig:MagnetIT}
    \end{figure}

\begin{figure}[H]
        \centering
        \includegraphics[width=1\columnwidth, keepaspectratio]{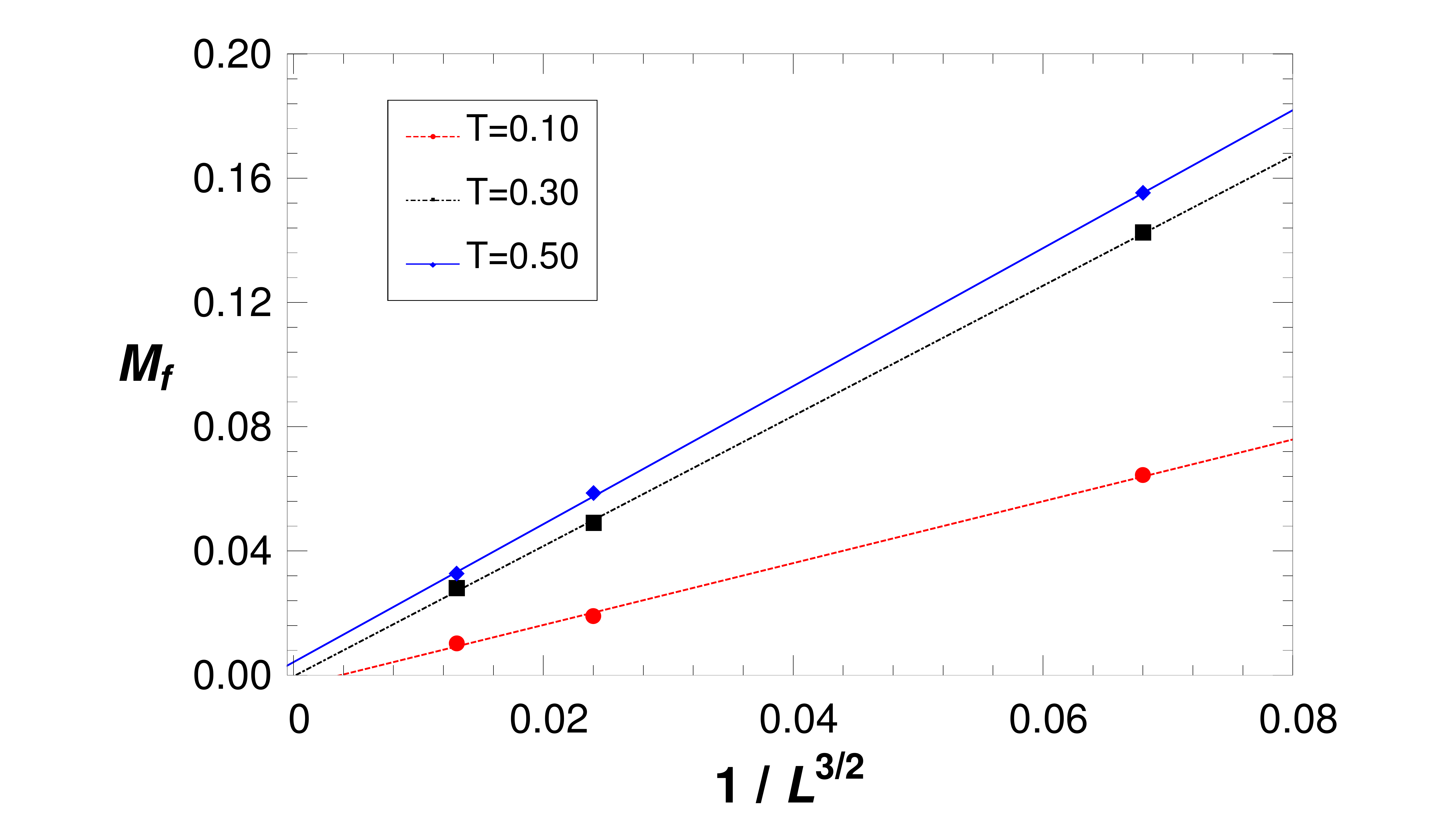}
        \caption[Total ferromagnetic magnetization vs the inverse power of lattice size]{Total ferromagnetic magnetization as a function of the inverse of lattice size to the power of $\frac{3}{2}$. This value is proportional to the inverse square root of the number of spins.}
        \label{fig:MagScaleInvL}
    \end{figure}

The total ferromagnetic magnetization is often used in classical spin systems as an order parameter, especially for those associated with dipolar coupling. In the present case, the ground state exhibits a zero net moment due to the alternating spin vectors along [111] axes. 
For the 2D kagome system, $M_{f}$ was shown to increase with decreasing temperature with a maximum value of $M_{f}=0.8700$ for a pure, single domain, ground state.\cite{holden2015}  Figure~\ref{fig:MagnetIT} shows MC results for $M_{f}$ decreasing with decreasing temperature, tending towards zero at low $T$.  This behaviour holds true for all lattice sizes, and $M_{f}$ diminishes rapidly with increasing $L$ at all $T$. Figure~\ref{fig:MagScaleInvL} shows the magnetization as a function of the inverse square root of number of spins at selected temperatures, showing a positive, linear scaling that grows with temperature. The magnetization approaches zero at all temperatures as lattice size increases while the slope of the fit decreases with temperature, as expected for an antiferromagnetic system. The ground state magnetization is expected to have a slope of zero (i.e. perfect antiferromagnetic order for all lattice sizes), which corresponds with this behaviour.
     
    %Paragraph 3 : Intro to sublattice magnetization and total sublattice magnetization & some results

 For the present system, as in the 2D case,\cite{holden2015} the sublattice magnetization serves as an order parameter associated with the phase transition to long range order. The value of this quantity in each individual sublattice of size $N/3$ is calculated as
    
    \begin{equation}\label{eq:SublatAntiF}
        M_{\gamma} = \dfrac{3}{N} \left| \sum_{i \subset \gamma}{S_{i} \left(-1\right)^{i}} \right|
    \end{equation}
    
    \noindent
    where $\gamma$ represents the subset of all spins belonging to any given sublattice, and $i$ is selected such that adjacent spins, $S_{i}$, carry an opposing sign (i.e. $i=n_{x}+n_{y}+n_{z}$ for lattice position $n=\left( n_{x}, n_{y}, n_{z} \right)$). The spins of each cubic sublattice are expected to approach perfect ordering ($M_{\gamma} =1$) at zero temperature. By taking a thermal average of each sublattice magnetization the total sublattice magnetization is calculated as
    
    \begin{equation}\label{eq:SublatTotalM}
        M_{t} = \dfrac{1}{3} \sum_{\gamma}{M_{\gamma}}
    \end{equation}
    
\begin{figure}[H]
        \centering
        \includegraphics[width=1\columnwidth, keepaspectratio]{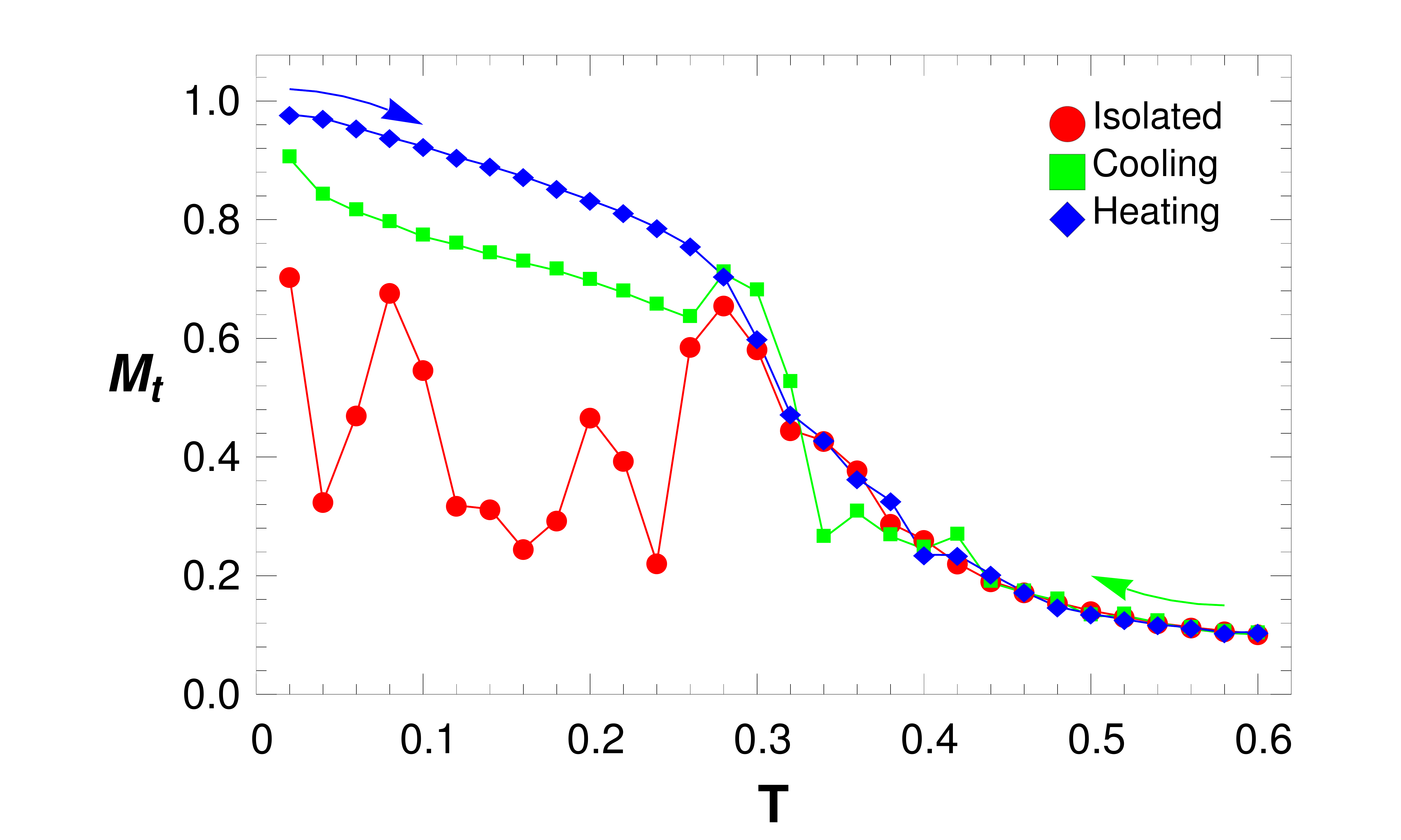}
        \caption[Total sublattice magnetization as a function of temperature for various simulations]{Total sublattice magnetization as a function of temperature for the three simulation types with $L=12$.}
        \label{fig:SublatTotal}
    \end{figure}

    \noindent
Figure~\ref{fig:SublatTotal} shows the results for the three different simulation types. It can be seen that in cooling simulations, some domains of ``frozen-in'' spins of a different orientation occur, as in the 2D case.\cite{holden2015} These domains reduce the total sublattice magnetization. In both cases of cooling and isolated temperature runs, for $T < T_{ne}$, for some temperature $T_{ne} \approx 0.30$, the system becomes non-ergodic and is no longer able to sample states effectively.

    % total sublattice magnetization
    
    % sublattice magnetization : C, H
    
    \begin{figure}[H]
        \centering
        \includegraphics[width=0.8\columnwidth, keepaspectratio]{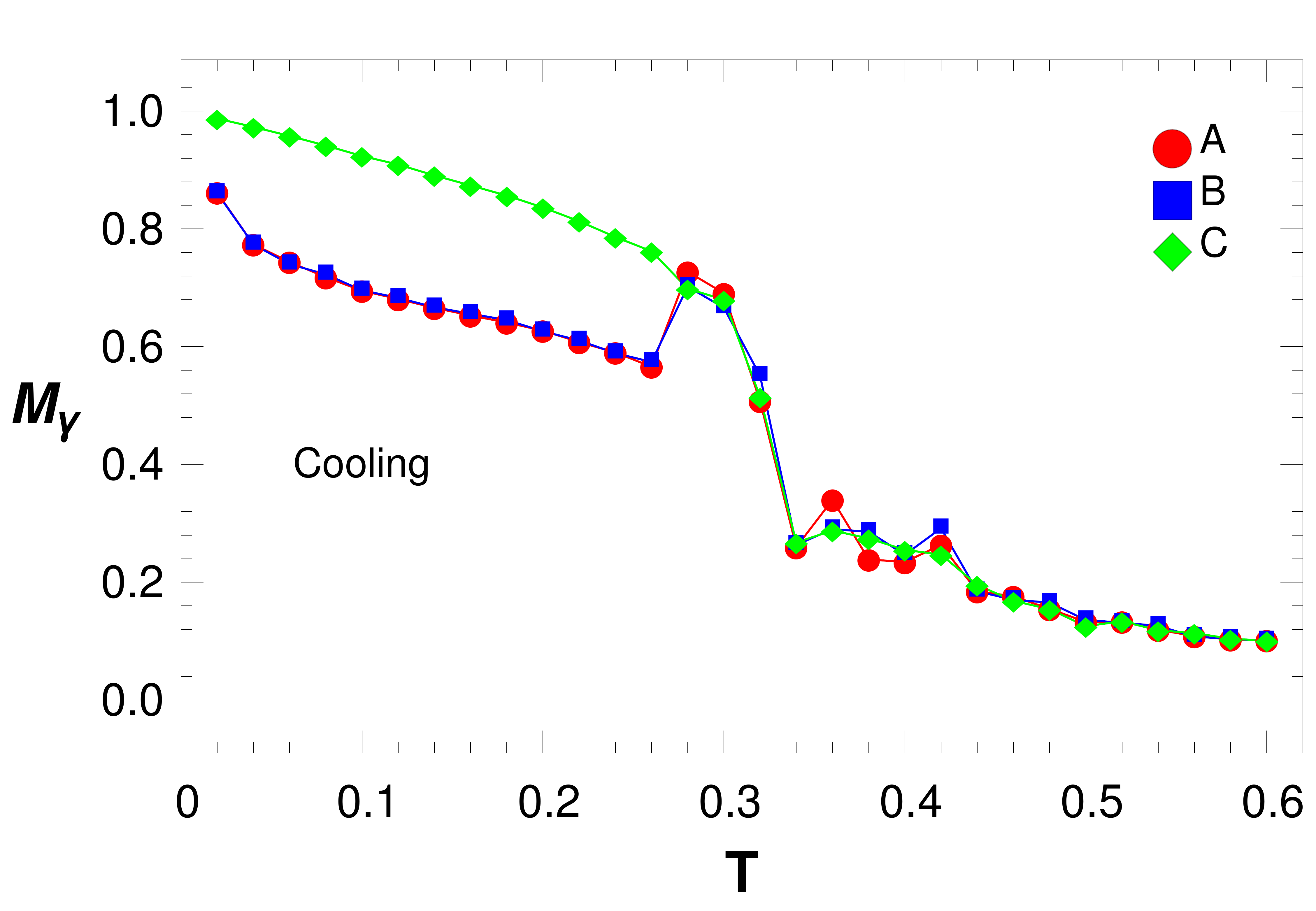}
        \includegraphics[width=0.8\columnwidth, keepaspectratio]{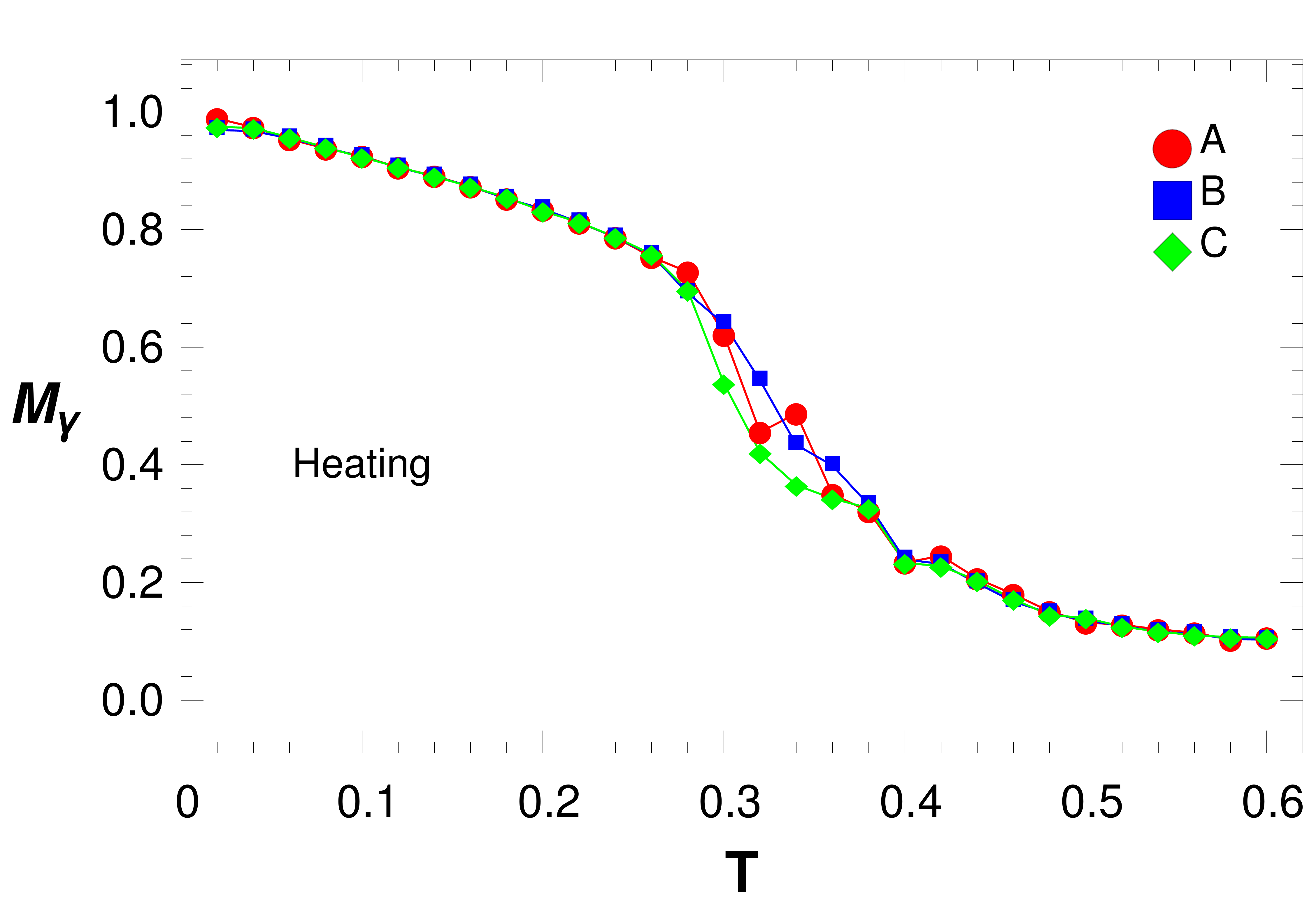}
        \caption[Sublattice antiferromagnetic magnetization of heating and cooling MC]{Individual sublattice magnetizations for $L=12$ cooling (top) and heating from a ground state (bottom) simulations. Frozen-in states occur in the $A$ and $B$ sublattices of the cooling simulations.}
        \label{fig:SublatL12}
    \end{figure}

\begin{figure}[H]
        \centering
        \includegraphics[width=1\columnwidth, keepaspectratio]{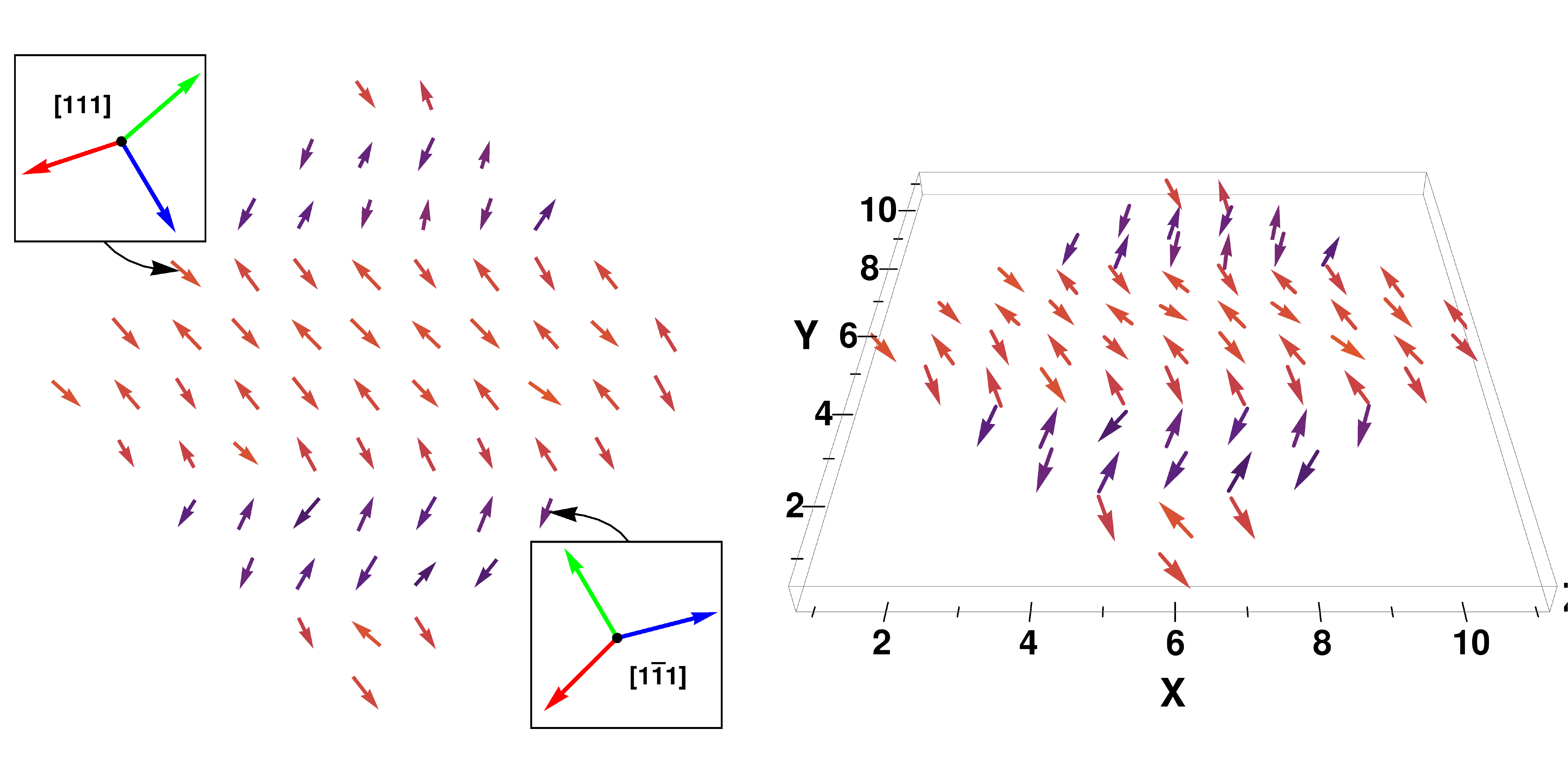}
\caption[Lattice cross section across the z axis of sublattice $A$ demonstrating frozen-in spin states.]{ A cross section of sublattice $A$  taken at $T=0.02$ across the z-axis. The 2D (left) representation only displays the angle $\phi$. The 3D (right) model presents the same cross section with $\theta$ included. Colors are determined by $\phi$, highlighting frozen-in domains.
Insets show $\vec{\mathbf{S}}_1$, $\vec{\mathbf{S}}_2$, and $\vec{\mathbf{S}}_3$ viewed along an indicated direction, and illustrate the geometric equivalence of the two domains.}
        \label{fig:CrossSectionZCool}
    \end{figure}

    %Paragraph 4 : discuss of frozen states
    The frozen-in states occur when it is no longer energetically favorable for single spins, even on the border of the domains, to change orientation.  Thermal averaging results for the three individual sublattice magnetizations are displayed in Fig.~\ref{fig:SublatL12} for cooling and heating simulations. Two of the cooled system sublattices do not uniformly approach unity. A depiction of frozen-in states in the $A$ sublattice is presented in the cross section of Fig.~\ref{fig:CrossSectionZCool} with a portion of differently-oriented spins highlighted. These frozen-in states occur in domains which are symmetrically equivalent on the lattice, spanning the entire lattice in one dimension (i.e.~individual plane), and appear any number of times. That is, these domains are related to each other by the crystal symmetry. When compared to a single-domain state, the energy differences at $T < 0.10$ are on the order of $\Delta E/spin \approx 10^{-3}$. Such a small energy difference leads to domain occurrence being commonplace in both isolated, cooling, and, on rare occasion, heating simulations.
    
    % frozen in states ; sublattice cross sections
    
    The ferromagnetic susceptibility, $\chi$, of the lattice is defined as the variance in ferromagnetic magnetization and is given by
    
    \begin{equation}\label{eq:Chi}
        \chi = \dfrac{1}{N_{\rm MCS}}\sum_{s}{\dfrac{(M_{f,s} - \left< M_{f}\right>)^{2}}{T}}.
    \end{equation}
    
    % \begin{equation}\label{eq:Chi} 
    %      \chi = \dfrac{(\left<M_{f}^{2}\right> - \left<M_{f}\right>^{2})}{T}
    % \end{equation}
    
\begin{figure}[H]
        \centering
        \includegraphics[width=1\columnwidth, keepaspectratio]{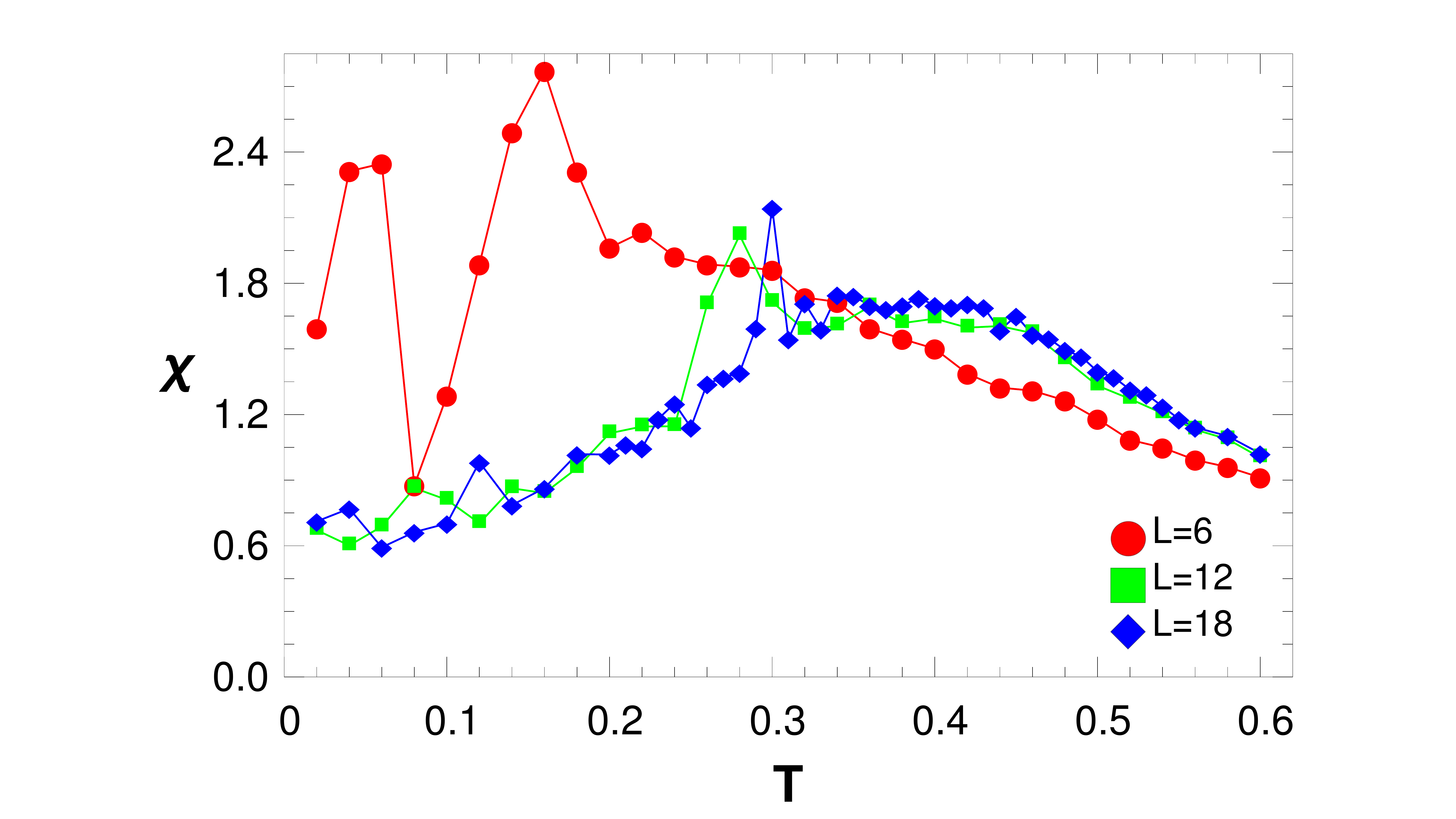}
        \caption[Ferromagnetic susceptibility vs temperature for varying lattice size at isolated temperatures.]{Ferromagnetic susceptibility ($\chi$) per spin for different lattice sizes $L$ for isolated temperature MC simulations.}
        \label{fig:ChiIT}
    \end{figure}

\begin{figure}[H]
        \centering
        \includegraphics[width=1\columnwidth, keepaspectratio]{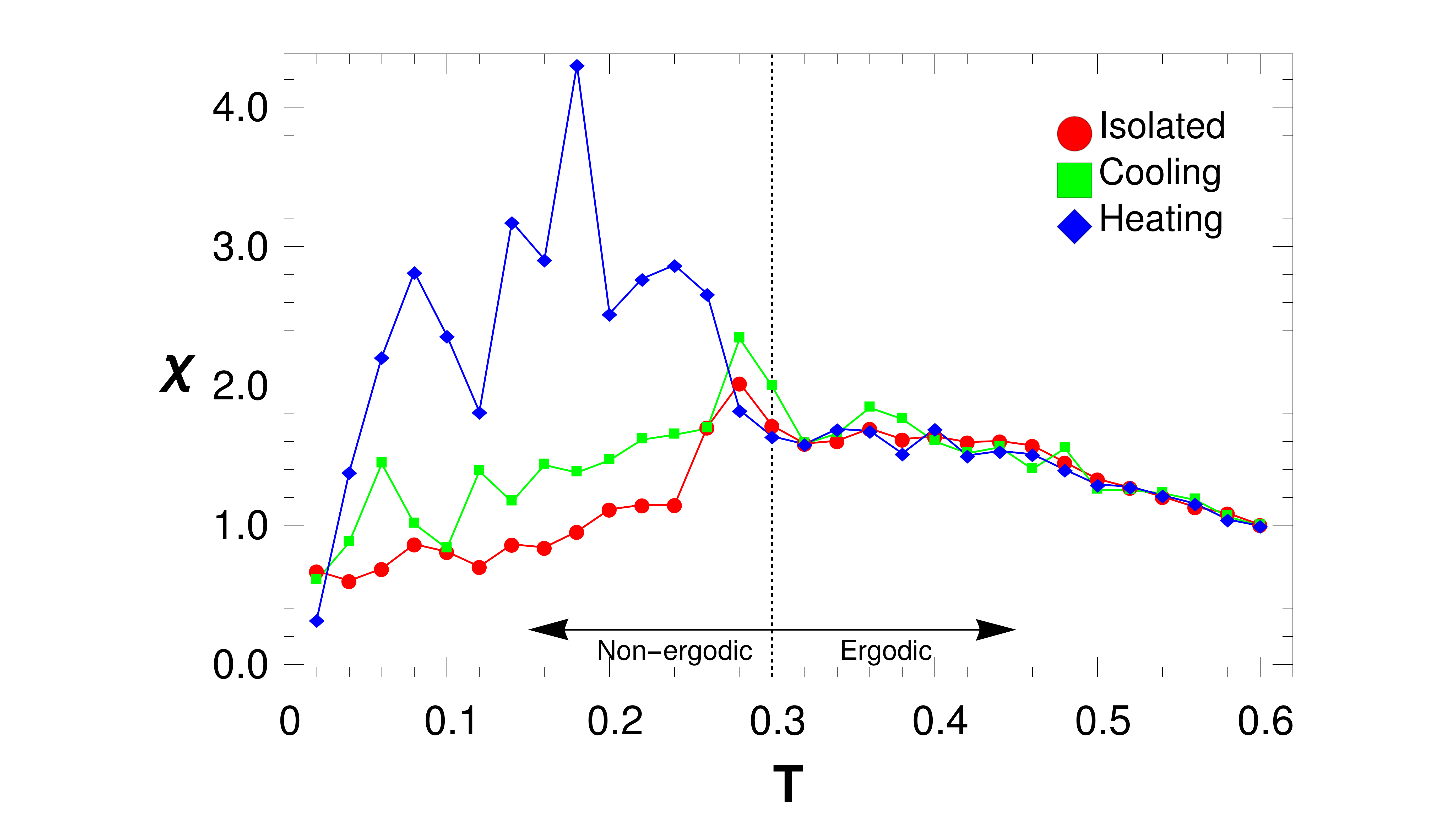}
        \caption[Ferromagnetic susceptibility for $L=12$]{Ferromagnetic susceptibility for $L=12$ from heating, cooling and isolated temperature simulations.}
        \label{fig:ChiL12}
    \end{figure}

    \noindent
Figure~\ref{fig:ChiIT} shows $\chi$ from isolated temperature simulations with varying lattice sizes. Figure~\ref{fig:ChiL12} shows $\chi$ for isolated temperature, and cooling and heating simulations at $L=12$ only. The majority of simulations produce a narrow, noisy peak at $T \approx 0.30$ followed by a weak, broad peak around $T=0.40$. The first, narrow peak occurs at the temperature at which the lattices become locked into their configurations, as seen in the total sublattice magnetization, is related the the non-ergodicity of the system below this temperature.  These per site values of the susceptibility do not appear to scale with respect to the system size, which is consistent with a lack of a ferromagnetic transition in the system.
    
    %Paragraph 6 : Discussion of the exceptions
    The obvious exceptions to this generalization are the $L=6$ isolated temperature runs where finite-size effects are strongest.   This may facilitate the sampling of more states at lower $T$, i.e., increase ergodicity. 
 The heating simulation results in Fig.~\ref{fig:ChiL12} at $L=12$ also show signifcant noise at lower $T$, as with the heating results in Fig.~\ref{fig:SublatTotal}, where spins remain locked into a ground state until the transition begins at $T \approx 0.30$. This behaviour is witnessed in all simulations, including the cooling simulations in which domains are not realized. Heating simulation results, in which the spin lattice quickly develops domains, align closely with the cooling simulation susceptibility suggesting the domains serve to lock spins into an orientation and reduce the susceptibility of the system.     
    
    % susceptibility : C, H, IT, L=12
    
Similarly, the antiferromagnetic susceptibility, $\chi_{a}$, can be defined in terms of the variance in the total sublattice magnetization defined in Eq.~\ref{eq:SublatTotalM}. This is given as
    
    \begin{equation}\label{eq:ChiA}  
        \chi_{a} = \dfrac{1}{N_{\rm MCS}}\sum_{s}{\dfrac{(M_{t,s} - \left< M_{t}\right>)^{2}}{T}}.
    \end{equation}
    
\begin{figure}[H]
        \centering
        \includegraphics[width=0.9\columnwidth, keepaspectratio]{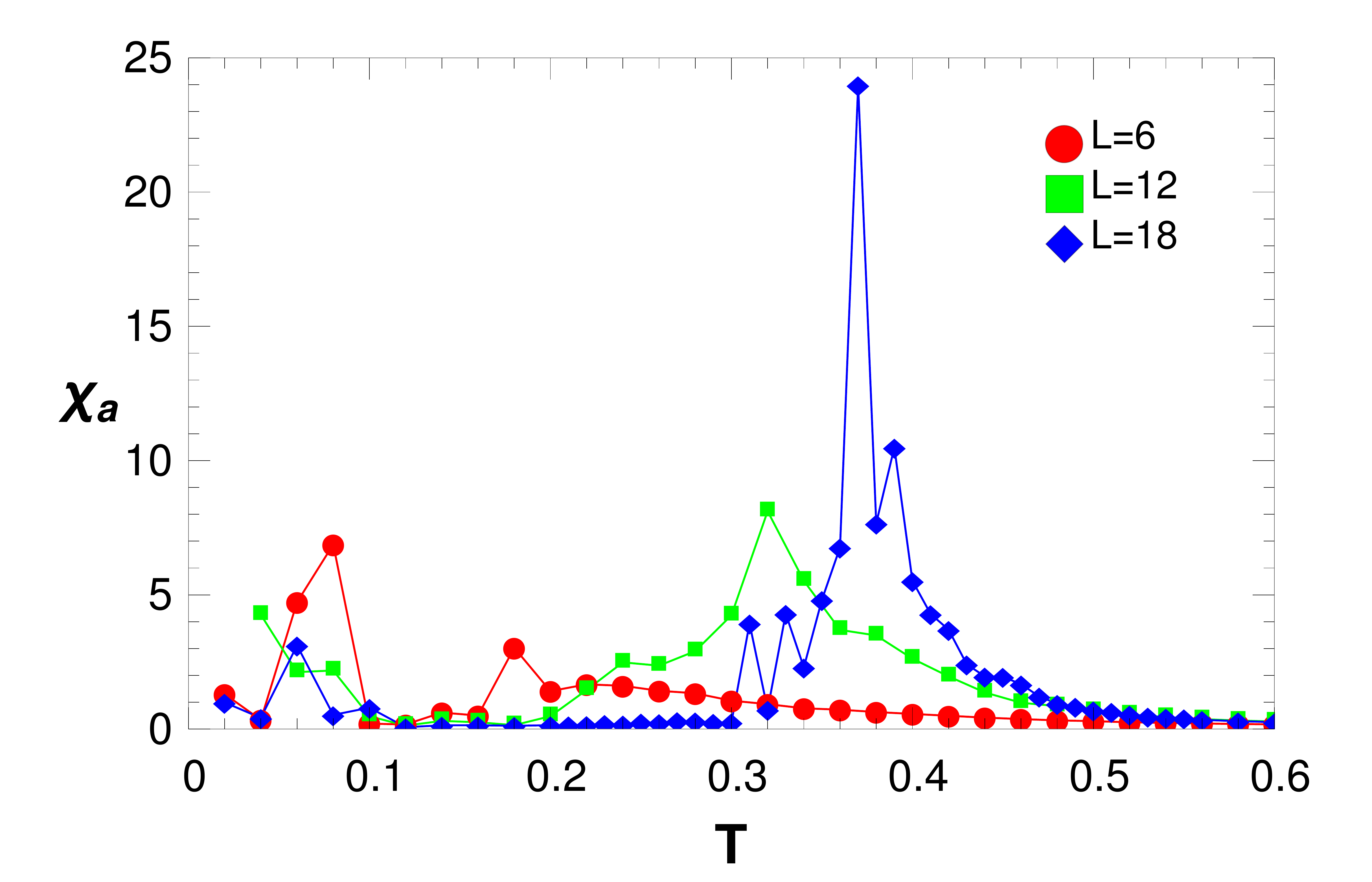}
        \caption[Antiferromagnetic susceptibility for different lattice sizes, at isolated temperatures]{Antiferromagnetic susceptibility for different lattice sizes from isolated temperature simulations.}
        \label{fig:ChiAIT}
    \end{figure}

    \noindent
Figure~\ref{fig:ChiAIT} shows results for isolated temperature simulations at $L$ = 6, 12, and 18. A single, sharp peak is apparent only for $L=18$, corresponding with the inflection point in the value of $M_{t}$, and peak in the specific heat, producing the same estimated critical temperature $T_{c}\simeq 0.38$.  Also note that for $L$ = 18, the results show reduced noise relative to the smaller lattice simulations below a temperature around 0.3, also consistent with the above discussion of lock-in states.
    
       % antiferromagnetic susceptibility
    
\section{Summary and Conclusions}\label{sec:summary}

This work explores an example of the interplay between the geometrical frustration of a 3D kagome lattice involving the antiferromagnetic and anisotropic character of the dipole interaction, as well as the long-range nature of  dipole coupling.  The ABC stacking of kagome planes gives rise to a truly 3D example of a kagome lattice, with eight NN sites, four in-plane and two connecting each adjacent plane. In contrast with the 2D case where discrete degeneracy of the three-sublattice ground state spin structures reflects the six-fold hexagonal anisotropy,\cite{maksy2015,holden2015,maksy2017} and the regular fcc lattice showing ferromagnetic order, the 3D fcc dipolar kagome lattice is shown here to exhibit a continuous degeneracy involving two parameters (the polar angles $\theta$ and $\phi$), similar to the simple cubic dipolar lattice.\cite{belobrov}  Certain regions in the $\theta$-$\phi$ plane are excluded from the domain of allowed ground states characterized by six sublattice spins, involving three spins around a triangle in the (111) plane, and three spins around the adjacent plane triangle pointing in opposite directions.  The total magnetization is thus zero, unlike the 2D case.

 This continuous degeneracy is removed by thermal fluctuations at low temperatures in an order-by-disorder process, and through magnetic field cycling, with both processes yielding planar states for which all six sublattice spins lie in a single plane. Thermal fluctuations result in the system approaching any one of a set of discrete ``node states'' ($\theta=\pi/2$, $\phi=\pi/4$ and odd multiples thereof).  Field cycling appears to pick out a set of discrete set of states along the locus of planar states, including node states.

%
%The resulting planar states have all six sublattice spins lying in a single plane, {\color{red} which has a continuously degenerate orientation characterized by a single parameter [thermal lifting of the continuous degeneracy selects a discrete set of node states, while the field seems to pick points along the planar curve]}.

The evolution of a degenerate ground state spin configuration with an applied magnetic field is dependent on the field direction, with M vs H curves showing multiple features for H along the [100] and [110] directions involving spin flop transitions of the sublattices. In the case of H along [111]  there is a transition at low fields from the non-planar state to a planar state and a second transition  at higher fields near saturation.  The behaviour of the energy under field recycling indicates that the non-planar states become metastable in the presence of an applied field. The detailed evolution of a state with increasing field is dependent on which of the ground states is used, but always returning to a planar state as the field decreases to zero.

%The evolution of a degenerate ground state spin configuration with an applied magentic field is dependent on the field direction, with M vs H curves showing features for H along [100] and [110] directions, but only a continuous rotation of spins in the case of H along [111] up to saturation. The detailed evolution of a state with increasing field is dependent on which of the ground states is used, but always returning to a planar state as the field decreases to zero.  

MC simulations show evidence for long-range order at $T \simeq 0.38$ and confirm the absence of a finite magnetization.  In addition, the MC results suggest that for $T < 0.3$, the system can lock-in to a mix of the degenerate ground state configurations and exhibit non-ergodic theromodynamics.    

The realization of such a purely dipolar magnetic system appears limited to artificial nanostructures\cite{artificial} but the relevance of the present work my be related to a number of atomic kagome-based structures which also exhibit exchange interactions.\cite{dun,hayashida}  Among AB$_3$ compounds where magnetic ions form the fcc kagome structure, IrMn$_3$ with strong AF exchange is the most well known for its technologically important role in spin valves.  As mentioned in Ref. [\onlinecite{holden2015}], dipole coupling may be expected to be important at the interface with an adjacent thin-film ferromagnet when IrMn$_3$ is used as an exchange bias material.Our results illustrate the importance of temperature and field dependent dipole energy barriers, relevant to the field cooling protocol required for setting the exchange bias field.\cite{ogrady} 
The present work may also be relevant to the magnetism in a sister compound, GeMn$_3$, which exhibits exchange-driven ferromagnetism where dipole coupling can also be important.\cite{gemn}

\section{Acknowledgements}\label{sec:acknowledgements}

This work was supported by the Natural Sciences and Engineering Research Council (NSERC) of Canada,
and the Compute Canada facilities of the Atlantic Computational Excellence network (ACEnet) and the Western Canada Research Grid (WestGrid).

\end{document}